\documentclass{aastex631}
\usepackage{longtable}
\usepackage{booktabs}  
\usepackage{amsmath}
\usepackage{graphicx}
\usepackage{booktabs}

\begin{document}

\title{Frequent optical flares in an M-type detached eclipsing binary TIC 349005337}

\correspondingauthor{Xiao Zhou}
\email{zhouxiaophy@ynao.ac.cn}

\author[0000-0002-0395-2205]{Xiao Zhou}
\affiliation{Yunnan Observatories, Chinese Academy of Sciences, Kunming 650216, China}
\affiliation{Key Laboratory of the Structure and Evolution of Celestial Objects, Chinese Academy of Sciences, Kunming 650216, China}
\affiliation{Center for Astronomical Mega-Science, Chinese Academy of Sciences, Beijing 100012, China}

\author{Liying Zhu}
\affiliation{Yunnan Observatories, Chinese Academy of Sciences, Kunming 650216, China}
\affiliation{Key Laboratory of the Structure and Evolution of Celestial Objects, Chinese Academy of Sciences, Kunming 650216, China}
\affiliation{Center for Astronomical Mega-Science, Chinese Academy of Sciences, Beijing 100012, China}

\author{Wei Hu}
\affiliation{Institute of Optoelectronic Technology, Lishui University, Lishui 323000, China}

\author{Gil Nachmani}
\affiliation{School of Physics and Astronomy, Faculty of Exact Sciences, Tel Aviv University, Tel Aviv 69978, Israel}

\author{Jia Zhang}
\affiliation{College of Physics and Optoelectronic Engineering, Leshan Normal University, Leshan 614000, China}

\author[0000-0002-9975-7833]{Mikhail Yu. Kovalev}
\affiliation{Yunnan Observatories, Chinese Academy of Sciences, Kunming 650216, China}
\affiliation{International Centre of Supernovae, Yunnan Key Laboratory, Kunming 650216, China}

\author{Ergang Zhao}
\affiliation{Yunnan Observatories, Chinese Academy of Sciences, Kunming 650216, China}
\affiliation{Key Laboratory of the Structure and Evolution of Celestial Objects, Chinese Academy of Sciences, Kunming 650216, China}
\affiliation{Center for Astronomical Mega-Science, Chinese Academy of Sciences, Beijing 100012, China}

\begin{abstract}
A comprehensive photometric and spectroscopic analysis of TIC 349005337, a highly active M-type detached eclipsing binary with an orbital period of 2.6079133 days, is presented utilizing high-cadence and long-term monitoring photometry from the Transiting Exoplanet Survey Satellite (TESS) and medium-resolution spectroscopy from the Large Sky Area Multi-Object Fiber Spectroscopic Telescope (LAMOST). By integrating radial velocity curves with light curves, we determined the fundamental physical parameters for both components. The system consists of an M-dwarf primary ($M_1 = 0.517 \pm 0.009 M_\odot$, $R_1 = 0.518 \pm 0.003 R_\odot$) and a fully convective secondary ($M_2 = 0.298 \pm 0.004 M_\odot$, $R_2 = 0.357 \pm 0.003 R_\odot$). A total of 12 high-energy stellar flares, with the most energetic event reaching $3.87 \times 10^{34}$ erg, is identified, confirming the system is in a state of magnetic hyper-activity. The frequent superflares suggest that intense magnetic activity and significant starspot coverage play a dominant role in altering the stellar atmospheric structure. As the secondary component falls below the fully convective boundary, these energetic flares provide critical empirical constraints for distributed dynamo models in the low-mass regime. Our results highlight TIC 349005337 as an ideal laboratory for investigating the interplay between binary dynamics, magnetic reconnection, and the resulting impact on the atmospheric stability of potential orbiting exoplanets.
\end{abstract}



\section{Introduction}         
\label{sect:intro}

M-dwarfs are the most numerically dominant stellar population in the Milky Way, serving as foundational benchmarks for understanding stellar structure and evolution in the low-mass regime \citep{2006AJ....132.2360H}. Eclipsing binaries (EBs) containing M-type components are particularly invaluable, as they allow for the direct, model-independent measurement of fundamental stellar parameters such as mass and radius \citep{2010A&ARv..18...67T}. However, a persistent and well-documented discrepancy exists between theoretical predictions and observational data: many M-dwarfs in binary systems exhibit ``radius inflation" \citep{1973A&A....26..437H,2012ApJ...757...42F} compared to standard stellar evolution models like MESA \citep{2011ApJS..192....3P}. This systematic bias is widely attributed to the presence of strong magnetic fields or high rotation rates, which may inhibit convective energy transport or enhance surface spot coverage \citep{2006Ap&SS.304...89R,2013ApJ...779..183F,2019MNRAS.482.5379G}. To resolve these theoretical bottlenecks, it is imperative to expand the sample of well-characterized M-type eclipsing binaries (MEBs) using high-precision photometry and multi-epoch spectroscopy. Such data provide the empirical constraints necessary to calibrate mass-radius-luminosity relations and refine our understanding of the interior physics of fully convective stars \citep{2011ASPC..448..505S,2024MNRAS.531.1765L}.

Stellar flares are transient, explosive manifestations of magnetic reconnection in stellar atmospheres, resulting in the rapid release of stored magnetic energy across the entire electromagnetic spectrum \citep{2002A&ARv..10..313P,2010ARA&A..48..241B}. In the era of high-precision space-based photometry and large-scale spectroscopic surveys, the study of these phenomena has evolved from individual case observations to robust statistical characterizations, offering profound insights into stellar physics and exoplanetary environments \citep{2016PhR...663....1S,2021ApJS..253...35T}. Recent data from missions such as TESS and Kepler have revealed the occurrence of ``superflares", events with bolometric energies reaching $10^{33}$ to $10^{38}$ erg, not only on young, rapidly rotating stars but also on older, solar-type stars \citep{2012Natur.485..478M,2020AJ....159...60G}. The mechanism driving flares in low-mass M-dwarfs remains a central question in stellar astrophysics. Unlike the Sun, stars with masses below approximately $0.35 M_\odot$ are fully convective and lack a tachocline—the interface layer between the radiative and convective zones. Despite the absence of this layer, fully convective stars exhibit intense magnetic activity and frequent flaring. Theoretical simulations and observations suggest that a ``distributed dynamo" or ``small-scale dynamo" can generate significant magnetic flux, which eventually reaches a saturation point in rapidly rotating systems \citep{2008ApJ...676.1262B,2016Natur.535..526W}. Flare observations in these systems provide a unique laboratory to test these non-solar dynamo theories and investigate how magnetic topology evolves without a tachocline. 

Motivated by these ongoing questions, the present paper investigates the flaring activity and fundamental properties of the M-type detached eclipsing binary system TIC 349005337, identified within the LAMOST and TESS datasets. The structure of this paper is arranged as follows: Section 2 describes the light curves observed by TESS and the orbital period correction performed based on the primary and secondary light minima. Section 3 details the processing of the LAMOST medium-resolution spectra, the fitting of the radial velocity curves, and the determination of the mass ratio. Section 4 presents the modeling of the light curves and the derivation of the fundamental physical parameters of the binary system. Section 5 focuses on the detection and characteristic analysis of stellar flare events observed in TIC 349005337. Finally, Section 6 provides a systematic summary of the work presented in this research.

\section{Photometric observations from TESS}
\label{sect:tess}

The Transiting Exoplanet Survey Satellite (TESS) was launched in April 2018. It was designed to discover exoplanets around bright, nearby stars using the transit method. Equipped with four wide-field cameras sensitive to red-visible and near-infrared light (600–1000 nm), TESS observes each of its sky sectors for about 27 days, achieving high precision data with cadences ranging from 30 minutes to 20 seconds. To date, TESS has identified over 6,000 exoplanet candidates, and has contributed to broader astrophysics by detecting stellar flares and supernovae. 

\begin{figure}[h]
\centering
\includegraphics[width=0.8\linewidth]{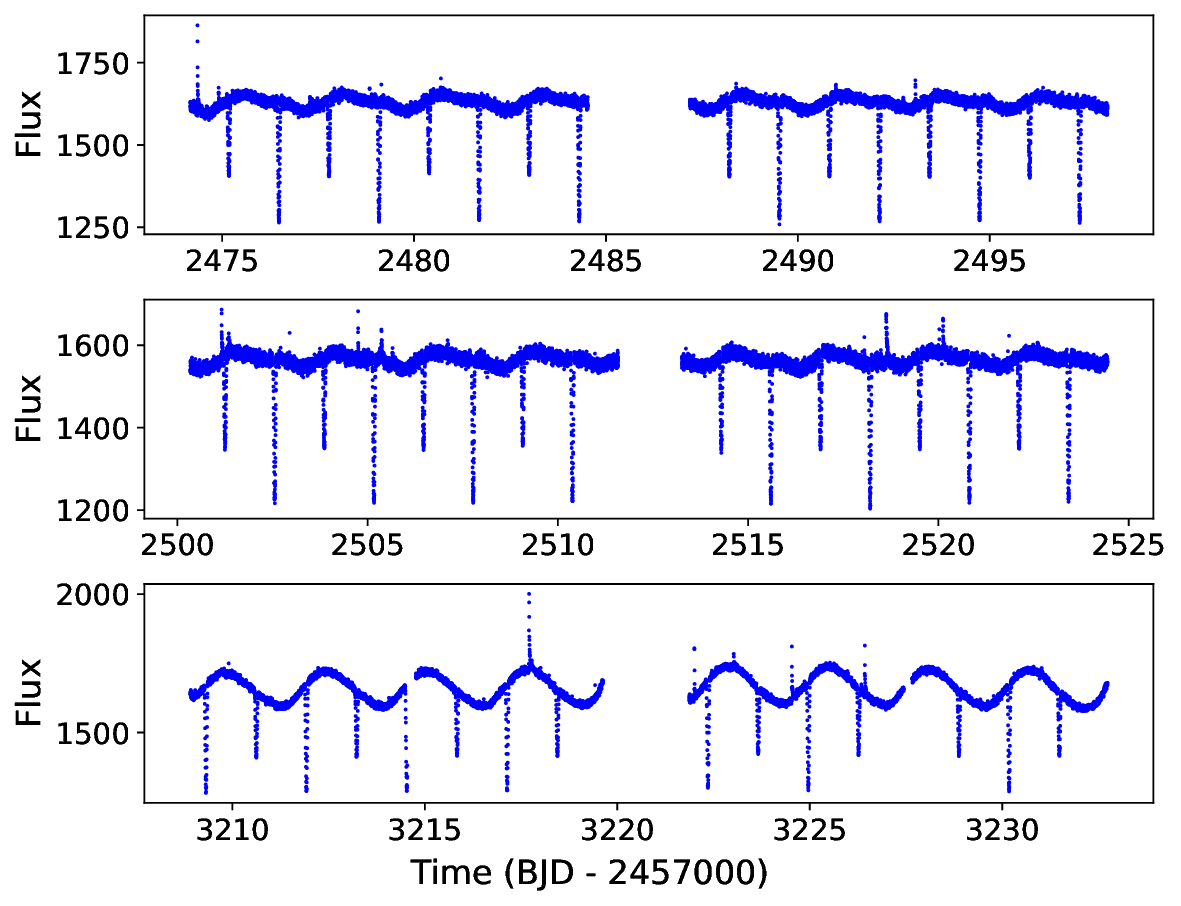}
\caption{The light curves from TESS Sector 43 (upper panel), Sector 44 (middle panel) and Sector 70 (lower panel).}
\label{tess-lc}
\end{figure}

The late-type star TIC 349005337 was observed by TESS in Sector 43, Sector 44, and Sector 70. The python package $lightkurve$ \citep{2018ascl.soft12013L} was used to get access to the observed light curves from the Mikulski Archive for Space Telescopes (MAST\footnote{\url{https://archive.stsci.edu}}). The light curves with 2 minutes' cadence were provided by the TESS science processing operations center (SPOC) in Sector 43 and Sector 44 \citep{2016SPIE.9913E..3EJ}. And the light curves with 200 seconds cadence were obtained from full-frame images (FFI) in Sector 70 \citep{2020RNAAS...4..201C}. Both Simple Aperture Photometry (SAP) flux and Pre-search Data Conditioning SAP (PDCSAP) flux are given by the data reduction pipeline. We prefer to use PDCSAP flux in our research since it is generally considered cleaner. PDCSAP flux is the SAP flux with long-term instrumental trends removed using cotrending basis vectors. The ultimately selected light curves are displayed in Fig. \ref{tess-lc}. It is obvious that many flares are observed in TIC 349005337.

To determine the eclipse timings and errors in Fig. \ref{tess-lc}, we defined a parabolic model and the MCMC hammer $emcee$ was applied. A total of 46 eclipse timings are determined and listed in Table \ref{Minima}, which include 22 primary light minima and 24 secondary light minima. The eclipse timings and their errors are listed in the first and third column, respectively. In the second column, the letter p refers to primary light minimum and the letter s refers to secondary light minimum. Then, the following equation:
\begin{equation}
Min.I(BJD) = 2459476.4789+2^{d}.6079133\times{E}. \label{linear-eaqution}
\end{equation}
is used to calculate the cycle numbers relative to the initial epoch and the corresponding values of the observed eclipse timings minus calculated eclipse timings (O - C values), which are listed in the fourth and fifth columns of Table \ref{Minima}. The initial epoch in Equation \ref{linear-eaqution} is determined with TESS data and the orbital period is adopted from \citet{2011AJ....141..166H}.

\begin{figure}[h]
\centering
\includegraphics[width=0.8\linewidth]{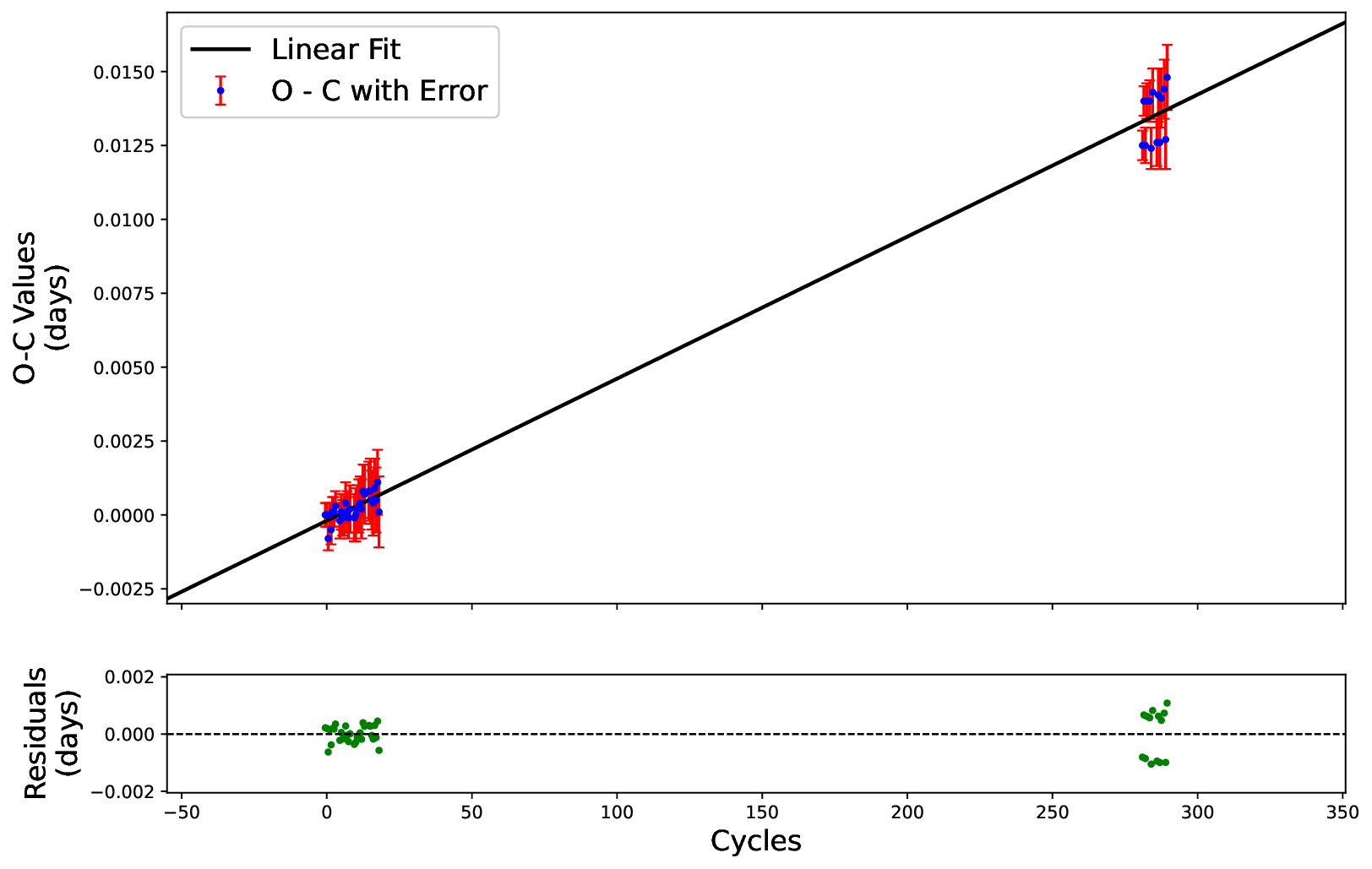}
\caption{Linear correction to the orbital period of TIC 349005337.}
\label{oc-fit}
\end{figure}

The O - C values and their errors are displayed in the upper panel of Fig. \ref{oc-fit}. Linear correction is applied to the orbital period of TIC 349005337, which is shown with a solid line in Fig. \ref{oc-fit}. The fitting residuals are displayed in the lower part of Fig. \ref{oc-fit}. The newly determined ephemeris of the eclipsing binary system is :
\begin{equation}
Min.I(BJD) = 2459476.4787(1)+2^{d}.6079614(4)\times{E}. \label{linear-eaqution-corrected}
\end{equation}

\begin{table}
\begin{center}
\caption{Eclipse timings and O - C values}\label{Minima}
\setlength{\tabcolsep}{4pt}
\begin{tabular}{llllrllllr}
\hline\hline
    BJD     &   p/s    &  Error    &   Cycle    &     $O - C$    &     BJD      &   p/s    &  Error    &   Cycle    &     $O - C$     \\
(2457000+)  &          &  (days)   &            &      (days)    &  (2457000+)  &          &  (days)   &            &      (days)     \\\hline
2475.1749   &    s     &  0.0004   &    -0.5	&     -0.0000	 &   2510.3825  &    p     &  0.0010   &    13.0	&      0.0007	  \\
2476.4789   &    p     &  0.0004   &     0.0	&      0.0000	 &   2514.2944  &    s     &  0.0010   &    14.5	&      0.0008	  \\
2477.7821   &    s     &  0.0004   &     0.5	&     -0.0008	 &   2515.5984  &    p     &  0.0011   &    15.0	&      0.0008	  \\
2479.0868   &    p     &  0.0004   &     1.0	&     -0.0000	 &   2516.9021  &    s     &  0.0010   &    15.5	&      0.0005	  \\
2480.3903   &    s     &  0.0005   &     1.5	&     -0.0005	 &   2518.2059  &    p     &  0.0011   &    16.0	&      0.0004	  \\
2481.6948   &    p     &  0.0005   &     2.0	&      0.0001	 &   2519.5104  &    s     &  0.0010   &    16.5	&      0.0009	  \\
2482.9988   &    s     &  0.0005   &     2.5	&      0.0001	 &   2520.8139  &    p     &  0.0011   &    17.0	&      0.0005	  \\
2484.3029   &    p     &  0.0005   &     3.0	&      0.0003	 &   2522.1185  &    s     &  0.0011   &    17.5	&      0.0011	  \\
2488.2143   &    s     &  0.0006   &     4.5	&     -0.0002	 &   2523.4214  &    p     &  0.0012   &    18.0	&      0.0001	  \\
2489.5186   &    p     &  0.0006   &     5.0	&      0.0001	 &   3209.3150  &    p     &  0.0005   &    281.0   &      0.0125	  \\
2490.8223   &    s     &  0.0006   &     5.5	&     -0.0001	 &   3210.6205  &    s     &  0.0005   &    281.5	&      0.0140	  \\
2492.1264   &    p     &  0.0007   &     6.0	&      0.0000	 &   3211.9230  &    p     &  0.0006   &    282.0	&      0.0125	  \\
2493.4307   &    s     &  0.0007   &     6.5	&      0.0004	 &   3213.2284  &    s     &  0.0006   &    282.5	&      0.0140	  \\
2494.7344   &    p     &  0.0007   &     7.0	&      0.0001	 &   3215.8363  &    s     &  0.0007   &    283.5	&      0.0140	  \\
2496.0381   &    s     &  0.0007   &     7.5	&     -0.0001	 &   3217.1387  &    p     &  0.0007   &    284.0	&      0.0124	  \\
2497.3424   &    p     &  0.0008   &     8.0	&      0.0002	 &   3218.4445  &    s     &  0.0008   &    284.5	&      0.0143	  \\
2501.2540   &    s     &  0.0008   &     9.5	&     -0.0001	 &   3222.3547  &    p     &  0.0008   &    286.0	&      0.0126	  \\
2502.5580   &    p     &  0.0009   &     10.0   &     -0.0000	 &   3223.6603  &    s     &  0.0009   &    286.5	&      0.0142	  \\
2503.8622   &    s     &  0.0008   &     10.5	&      0.0002	 &   3224.9626  &    p     &  0.0009   &    287.0	&      0.0126	  \\
2505.1662   &    p     &  0.0009   &     11.0	&      0.0003	 &   3226.2681  &    s     &  0.0010   &    287.5	&      0.0141	  \\
2506.4703   &    s     &  0.0009   &     11.5	&      0.0004	 &   3228.8763  &    s     &  0.0010   &    288.5	&      0.0144	  \\
2507.7741   &    p     &  0.0010   &     12.0	&      0.0002	 &   3230.1785  &    p     &  0.0010   &    289.0	&      0.0127	  \\
2509.0786   &    s     &  0.0009   &     12.5	&      0.0008	 &   3231.4846  &    s     &  0.0011   &    289.5	&      0.0148     \\         
\hline
\hline
\end{tabular}
\end{center}
\end{table}

\section{Medium-Resolution Spectra from LAMOST}
\label{sect:lamost}
The Large Sky Area Multi-Object Fiber Spectroscopic Telescope (LAMOST), also known as the Guo Shoujing Telescope, is a 4-meter reflective Schmidt telescope with 4000 fibers equipped on its focal plane of 5-degree's field of view, located at the Xinglong Observatory, National Astronomical Observatories of the Chinese Academy of Sciences \citep{2012RAA....12.1197C,2012RAA....12..723Z}. In September 2018, LAMOST started the medium-resolution spectroscopic survey (MRS) with $R \sim 7500$ and a limiting magnitude of around $G = 15$ mag, and the blue cameras cover the wavelength range from 4950 \AA{} to 5350 \AA, while the red cameras cover the wavelength range from 6300 \AA{} to 6800 \AA{} \citep{2020arXiv200507210L}. 

\subsection{Radial Velocities of the Primary Star}

TIC 349005337 was observed by LAMOST-MRS from November 28, 2018 to February 16, 2022. A total of 50 blue-arm spectra and 53 red-arm spectra were observed during this time in LAMOST-MRS DR12. The cross-correlation function (CCF) technique is used to measure the radial velocities. We set the stellar atmospheric parameters for the template spectrum to an effective temperature of $T_{\text{eff}}$ = 3500 K, a surface gravity of $\log g$ = 4.24, and a metallicity of [Fe/H] = 0. We then generated a theoretical spectrum using iSpec\citep{2014A&A...569A.111B} to serve as the CCF template. In CCF fitting, the wavelength ranges selected at the blue and red ends are 4890 Å to 5370 Å and 6260 Å to 6830 Å, respectively. Every observational spectrum was cross-correlated against a template spectrum over a selected wavelength range following the method of \citet{1979AJ.....84.1511T}. The resulting CCFs are fitted with Gaussian profiles, and the radial velocities (RVs) are determined from the positions of the CCF peaks. For TIC 349005337, only one component is visible in their CCF profiles, which means that it's a single-lined spectroscopic binary system. All measured radial velocities and their corresponding orbital phases are listed in Table \ref{RV-CCF}, and the orbital phases are calculated with Equation \ref{linear-eaqution-corrected}.

\begin{table}
\begin{center}
\caption{Measured radial velocities through the CCF method.}\label{RV-CCF}
\setlength{\tabcolsep}{4pt}
\begin{tabular}{llllll}
\hline\hline
Phase     &       RV1     &       Error  &     Phase   &      RV1      &      Error     \\  
          &     (km/s)    &      (km/s)  &             &     (km/s)    &      (km/s)     \\\hline
0.0760    &     -4.855    &      0.404   &   0.8659    &      69.121   &      0.575       \\
0.0824    &     -2.685    &      0.460   &   0.8723    &      67.606   &      1.045       \\
0.0885    &     -0.675    &      0.624   &   0.9490    &      44.814   &      0.836       \\
0.1833    &     -19.114   &      1.375   &   0.9525    &      43.255   &      0.354       \\
0.1894    &     -18.761   &      1.174   &   0.9551    &      44.669   &      0.576       \\
0.1955    &     -20.914   &      0.693   &   0.9588    &      43.020   &      0.360       \\
0.2019    &     -22.205   &      2.929   &   0.9615    &      39.043   &      0.446       \\
0.5198    &     36.563    &      0.273   &   0.9650    &      41.483   &      0.198       \\
0.5259    &     38.411    &      0.295   &   0.9714    &      38.741   &      1.534       \\
0.5320    &     40.356    &      0.294   &   0.9775    &      35.763   &      0.640       \\
0.5469    &     44.681    &      0.314   &   0.9836    &      34.976   &      0.623       \\
0.5531    &     47.586    &      0.422   &   0.9900    &      31.939   &      0.345       \\
0.7000    &     81.302    &      2.638   &   0.9961    &      30.314   &      0.304        \\
0.7059    &     79.726    &      1.387   &             &               &                   \\
\hline
\hline
\end{tabular}
\end{center}
\end{table}

\subsection{The Double-peaked H$\alpha$ Emission Profile}
\label{subsec:q_rv}
\begin{figure}[h]
\centering
\includegraphics[width=0.8\linewidth]{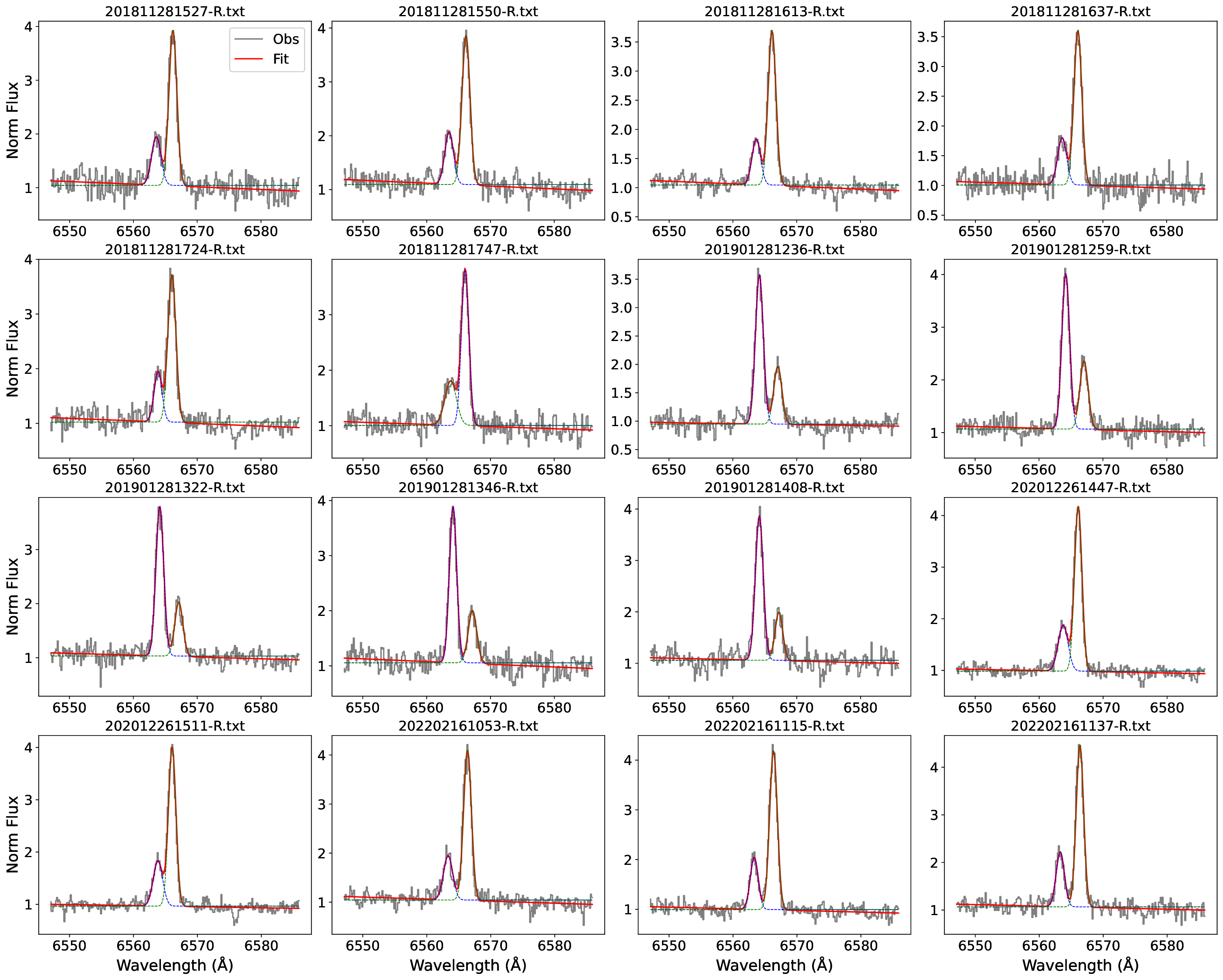}
\caption{Double-Gaussian fitting of H$\alpha$ emission lines for the LAMOST-MRS. Gray stepped lines represent the observed data, while solid red lines indicate the best-fit double-Gaussian models. Individual Gaussian components (main and sub-peaks) are shown in blue and green dashed lines, respectively.}
\label{H-alpha-fit}
\end{figure}

The double-peaked H$\alpha$ emission structure was observed in the red-arm spectra of LAMOST-MRS on 28 November 2018, 28 January 2019, 26 December 2020, and 16 February 2022. The radial velocities of the binary components were determined through a multi-step spectral line profile analysis centered on the H$\alpha$ emission line. First, the local continuum of each spectrum was normalized using a linear baseline fit to the line-free regions (6540–6550 Å and 6580–6590 Å). To resolve the overlapping contributions from the primary and secondary stars, we employed a double-Gaussian decomposition model integrated with a linear background term, defined as $f(\lambda) = \sum_{i=1}^{2} A_i \exp\left(-\frac{(\lambda - \mu_i)^2}{2\sigma_i^2}\right) + (b + k\lambda)$. The model parameters were optimized using a non-linear least-squares Levenberg-Marquardt algorithm. The centroid wavelengths ($\mu_1, \mu_2$) derived from the best-fit profiles were then converted into radial velocities via the Doppler shift formula, $v_r = c \cdot (\lambda_{obs} - \lambda_0) / \lambda_0$, where $\lambda_0$ is the vacuum wavelength of H$\alpha$ (6564.61 Å). Uncertainties were estimated from the diagonal elements of the covariance matrix, accounting for both photon noise and fitting residuals. The Double-peaked H$\alpha$ Emission Profile and corresponding fitting results are displayed in Fig. \ref{H-alpha-fit}. The measured radial velocities are listed in Table \ref{RV-H-alpha}, and the orbital phases are also calculated with Equation \ref{linear-eaqution-corrected}.

\begin{table}
\begin{center}
\caption{Measured radial velocities through H$\alpha$ emission lines.}\label{RV-H-alpha}
\setlength{\tabcolsep}{4pt}
\begin{tabular}{llllll}
\hline\hline
Phase    &       RV1     &       Error      &        RV2     &      Error    &    Observed date      \\  
         &     (km/s)    &      (km/s)      &       (km/s)   &      (km/s)   &                         \\\hline
0.8433   &    72.368     &	   2.827        &     -46.583 	 &    9.854      &   28 November 2018       \\
0.8494   & 	  70.932     &     1.077        &     -50.432 	 &    3.184      &   28 November 2018      \\
0.8556   & 	  69.613     &	   0.679        &     -43.360	 &    2.408      &   28 November 2018       \\
0.8619   & 	  66.808     &	   2.982        &     -45.869    &   10.636      &   28 November 2018        \\
0.8745   & 	  65.211   	 &     0.934        &     -34.457 	 &    2.749      &   28 November 2018        \\
0.8806   &    64.558     &	   1.094        &     -39.334    &    4.899      &   28 November 2018         \\
0.1869   &   -20.619     &	   1.611        &     111.083	 &    4.238      &   28 January 2019         \\
0.1930   &   -19.344     &	   0.709        &     110.794	 &    1.666      &   28 January 2019         \\
0.1992   &   -21.141     &	   0.687        &     114.783	 &    1.920      &   28 January 2019         \\
0.2056   &   -21.740     &	   0.775        &     117.169	 &    2.465      &   28 January 2019         \\
0.2114   &   -21.277     &	   0.794        &     118.543 	 &    2.335	     &   28 January 2019        \\
0.8669   & 	  68.281     &	   0.546        &     -39.683 	 &    2.247      &   26 December 2020         \\
0.8733   & 	  66.113     &	   0.548        &     -35.841	 &    2.162      &   26 December 2020         \\
0.6995   & 	  80.076     &	   0.636        &     -58.707 	 &    2.366      &   16 February 2022         \\
0.7053   & 	  78.385 	 &     0.528        &     -59.268	 &    1.627      &   16 February 2022        \\
0.7112   & 	  79.991     &	   1.230        &     -60.516 	 &    3.831      &   16 February 2022        \\
\hline       
\hline
\end{tabular}
\end{center}
\end{table}

\begin{figure}[h]
\centering
\includegraphics[width=0.8\linewidth]{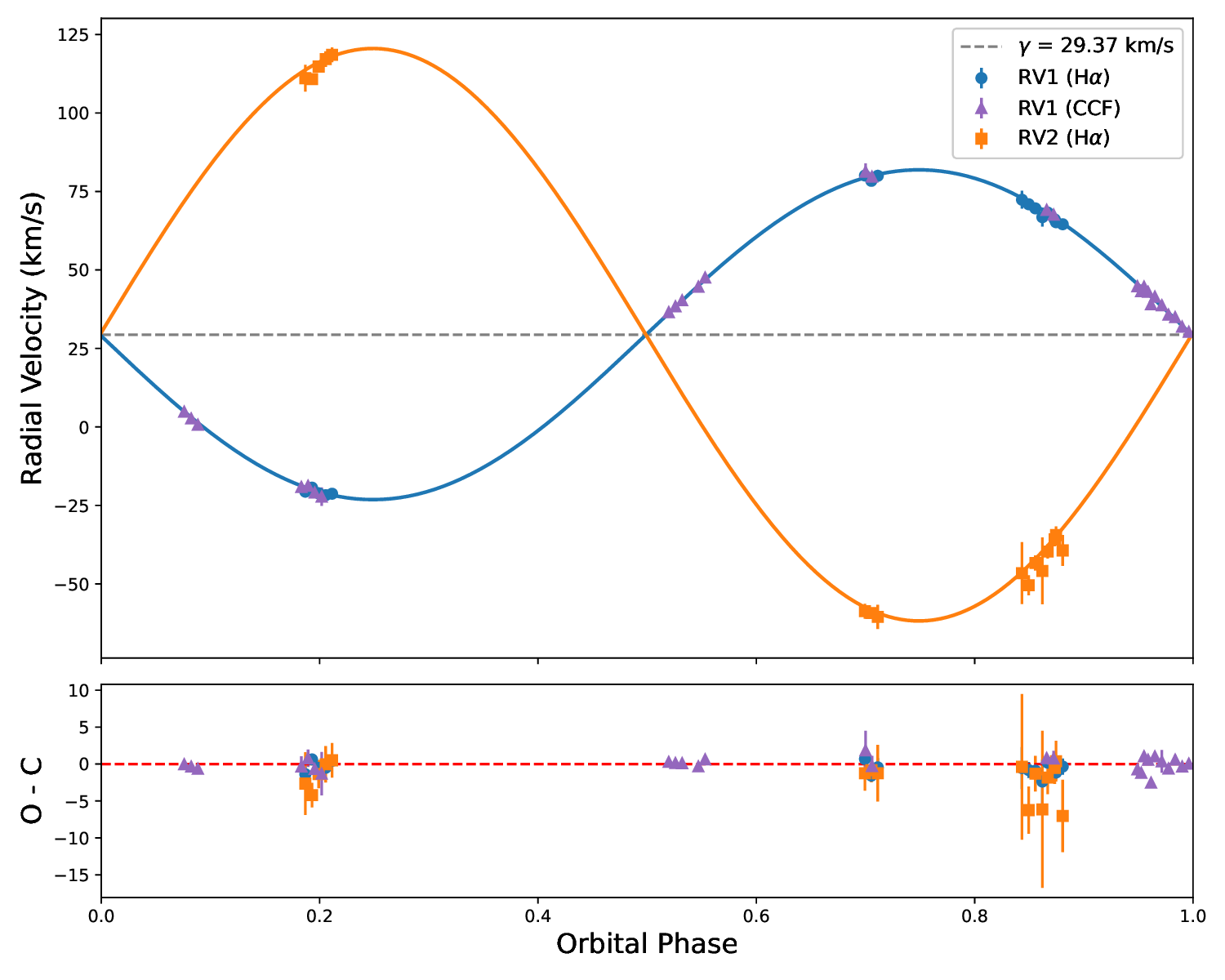}
\caption{Observed and fitted RV curves of TIC 349005337. Top panel: Observed radial velocities of the primary (blue circles and purple triangles) and secondary (orange squares) stars as a function of orbital phase. The purple triangles represent RV measurements derived from the Cross-Correlation Function (CCF) method, while other symbols denote H$\alpha$ line profile fitting results. Solid lines represent the best-fit circular Keplerian model derived from MCMC analysis, and the horizontal dashed line indicates the systemic velocity ($\gamma$). Bottom panel: Residuals ($O - C$) of the RV measurements relative to the best-fit model.}
\label{rv-fit}
\end{figure}

As shown in the upper panel of Fig. \ref{rv-fit}, the purple triangles represent radial velocities of the primary star derived from the Cross-Correlation Function (CCF) method. The blue filled circles represent radial velocities of the primary star derived from the H$\alpha$ emission lines, and the orange squares represent radial velocities of the secondary star derived from the H$\alpha$ emission lines. As illustrated in Fig. \ref{rv-fit}, the radial velocities of the primary star derived from the CCF method are consistent with those from H$\alpha$ line fitting. Consequently, we merged the two datasets to fit the RV curves and derive the mass ratio of the binary system.

\begin{figure}[h]
\centering
\includegraphics[width=0.8\linewidth]{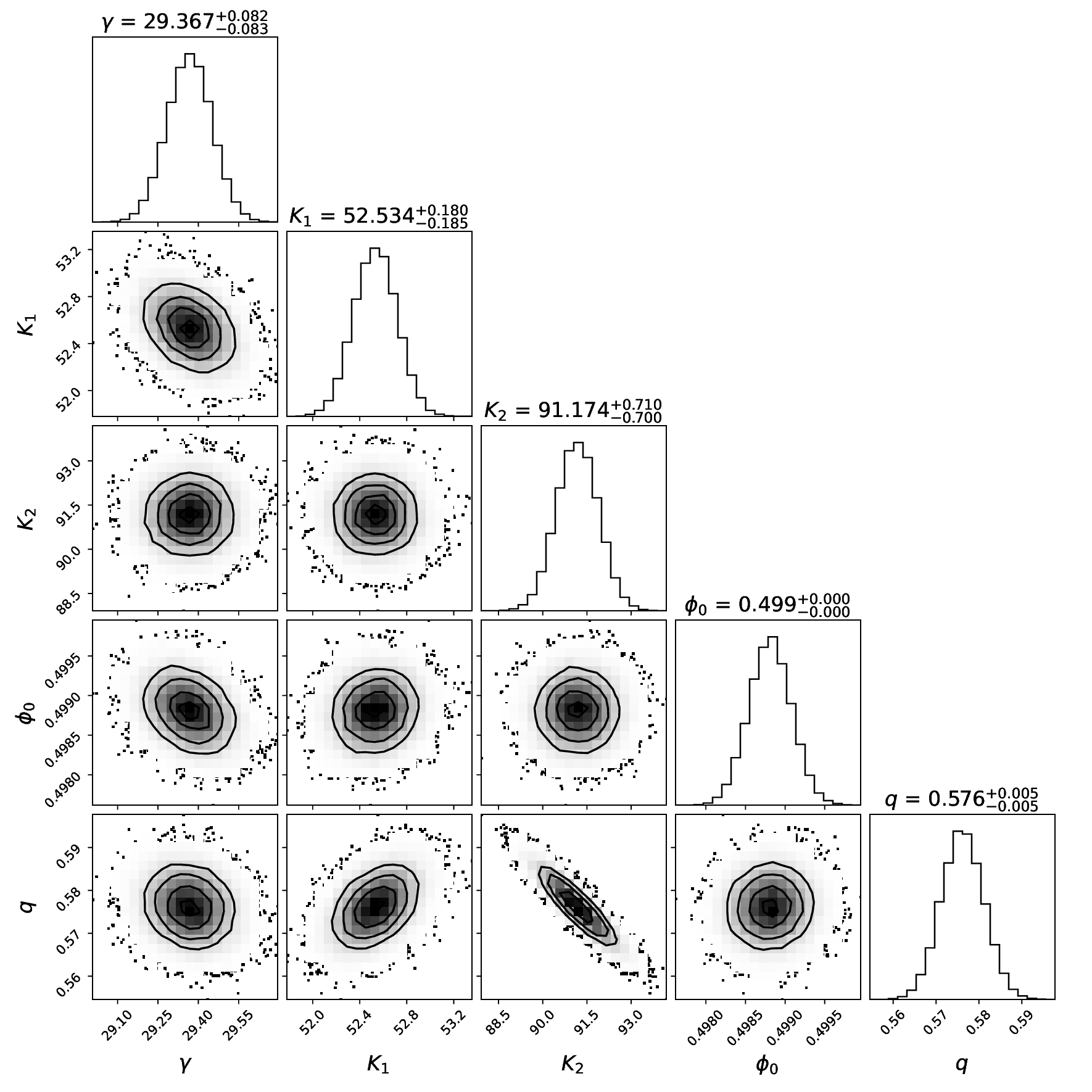}
\caption{Posterior distributions of the fitted parameters obtained from the MCMC analysis. Corner plot showing the one- and two-dimensional projections of the posterior probability distributions for the fitted parameters: systemic velocity ($\gamma$), semi-amplitudes ($K_1, K_2$), phase offset ($\phi_0$), and the derived mass ratio ($q$). The contour maps illustrate the correlations between different parameters, demonstrating a well-converged MCMC sampling process.}
\label{rv-corner}
\end{figure}

The measured radial velocities are fitted with a circular orbit:
\begin{equation}
RV1(\phi) = \gamma + K_1 \sin[2\pi(\phi - \phi_0)], \label{RV1}
\end{equation}
and 
\begin{equation}
RV2(\phi) = \gamma - K_2 \sin[2\pi(\phi - \phi_0)], \label{RV2}
\end{equation}

where $\gamma$ is the systemic velocity, $K_1$ and $K_2$ are the semi-amplitude, and $\phi_0$ is the phase zero-point. The fitting was performed using a Markov Chain Monte Carlo (MCMC) approach implemented with the emcee package \citep{2013PASP..125..306F}. The determined parameters are : $\gamma$ = $29.37^{+0.08}_{-0.08}$ km/s, $K_1$= $52.5^{+0.2}_{-0.2}$ km/s, $K_2$= $91.2^{+0.7}_{-0.7}$ km/s, $\phi_0$ = $0.4988^{+0.0003}_{-0.0003}$, and mass ratio $q$ = $0.576^{+0.005}_{-0.005}$. The observed RV data and the fitted RV curve are displayed in Fig. \ref{rv-fit}. The resulting posterior distributions are presented in the corner plot (Fig. \ref{rv-corner}). The one-dimensional distributions along the diagonal provide the median values and credible intervals for each parameter, while the two-dimensional distributions in the off-diagonal panels reveal potential correlations between parameters.

For a double-lined spectroscopic binary (SB2) with a circular orbit, the minimum mass of the component stars can be derived from the orbital period ($P$), semi-amplitudes $K_1$ and $K_2$:
\begin{equation}
M_1 \sin^3 i = \frac{P (K_1 + K_2)^2 K_2}{2\pi G}
\end{equation}
\begin{equation}
M_2 \sin^3 i = \frac{P (K_1 + K_2)^2 K_1}{2\pi G}
\end{equation}
Thus, the minimum mass of the primary and secondary stars are calculated to be $M_1 \sin^3 i$ = $0.509^{+0.009}_{-0.009}$ $M_\odot$ and $M_2 \sin^3 i$ = $0.293^{+0.003}_{-0.003}$ $M_\odot$

\subsection{Spectral disentangling}
\label{subsec:q_spd}
\begin{figure}[h]
\centering
\includegraphics[width=0.8\linewidth]{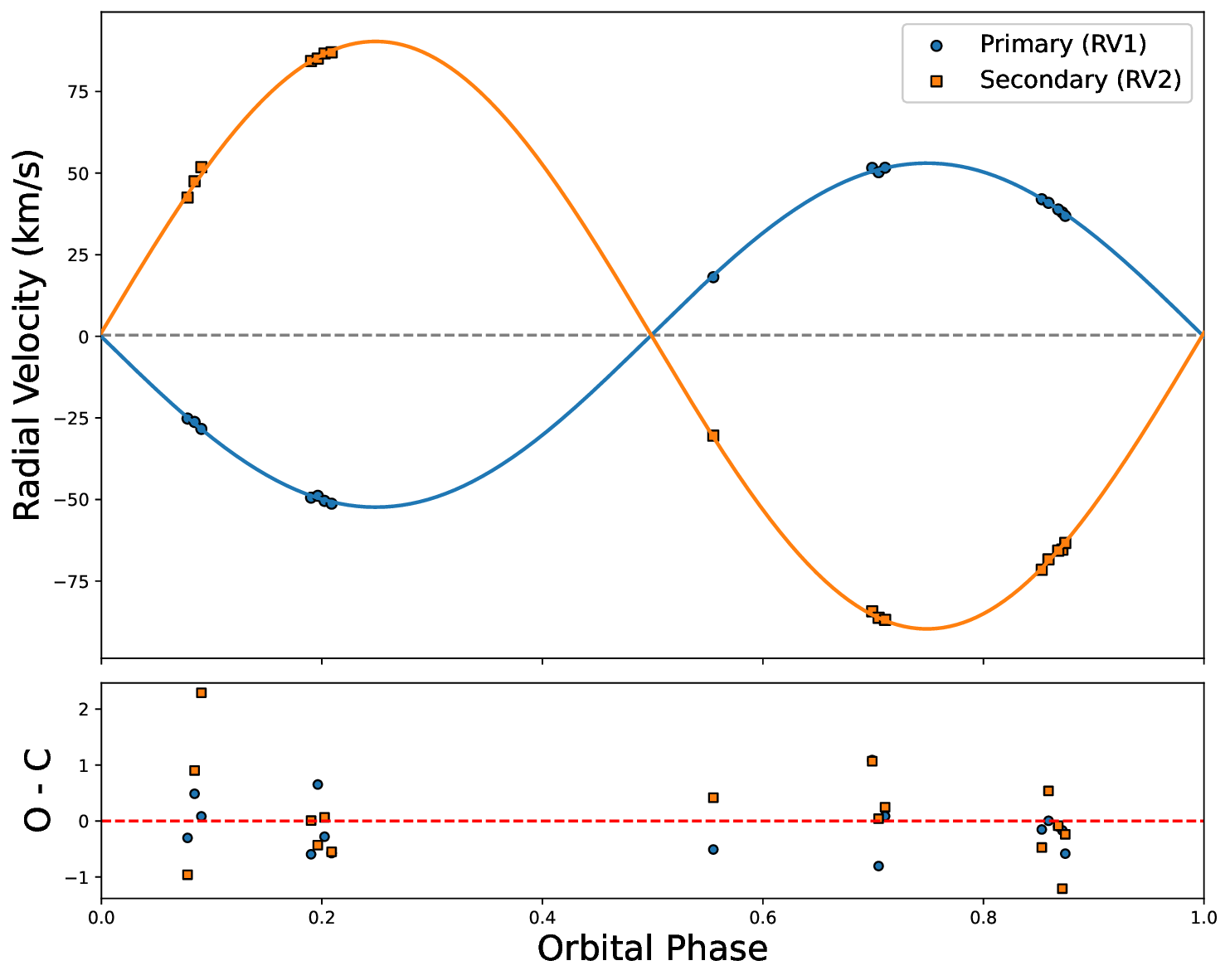}
\caption{Radial velocity curves and the best-fitting orbital model derived from spectral disentangling. The upper panel displays the measured radial velocities of the primary (blue circles, RV1) and secondary (orange squares, RV2) components. The solid lines represent the best-fitting sinusoidal models for both stars. The lower panel shows the corresponding residuals ($O - C$) for both components.}
\label{rv-spd}
\end{figure}

Since the CCF method only resolved the radial velocity for one of the two components, and considering that the mass ratio is an extremely critical physical parameter, we performed a re-analysis of the data using spectral disentangling. Previously this technique was successfully applied to LAMOST medium-resolution spectra in \citet{j0647,sb2wilson,2026arXiv260420314K}. We used Fourier disentangling code {\sc KOREL} \citep{korel,korelman} through web interface \url{https://stel.asu.cas.cz/vo-korel} \citep{vokorel}.  As illustrated in Table \ref{tab:rv_phase_spd} and Fig. \ref{rv-spd}, {\sc KOREL} successfully disentangled two spectral components and measured the radial velocities for them using 16 red arm LAMOST medium-resolution spectra, selected with signal-to-noise ratio S/N>5 and out of eclipse phases. We allow code to optimize only $t_0=59475.97784$ (BMJD) d, $K_1=52.83$ km/s and $q=0.5878$, while other parameters were fixed during the {\sc KOREL} run. 

To quantitatively characterize the orbital dynamics of the binary system, we fit the measured radial velocity data points obtained for both components. From the resulting best-fit orbital solution, we successfully determined the radial velocity semi-amplitudes, denoted as $K_1$= $52.7^{+0.3}_{-0.3}$ km/s and $K_2$= $90.0^{+0.3}_{-0.3}$ km/s, for the primary and secondary stars, respectively. Since the mass ratio of a spectroscopic binary is inversely proportional to the ratio of its velocity semi-amplitudes ($q = M_2/M_1 = K_1/K_2$), the mass ratio of TIC 349005337 is determined to be $q$ = $0.586^{+0.004}_{-0.004}$. These precisely derived velocity amplitudes allowed us to directly calculate the system's mass ratio without relying on any model-dependent assumptions. This process provides a robust observational constraint on the fundamental physical properties of the components. The minimum mass of the primary and secondary stars are calculated to be $M_1 \sin^3 i$ = $0.495^{+0.004}_{-0.004}$ $M_\odot$ and $M_2 \sin^3 i$ = $0.290^{+0.003}_{-0.003}$ $M_\odot$.

\begin{figure}[h]
\centering
\includegraphics[width=0.8\linewidth]{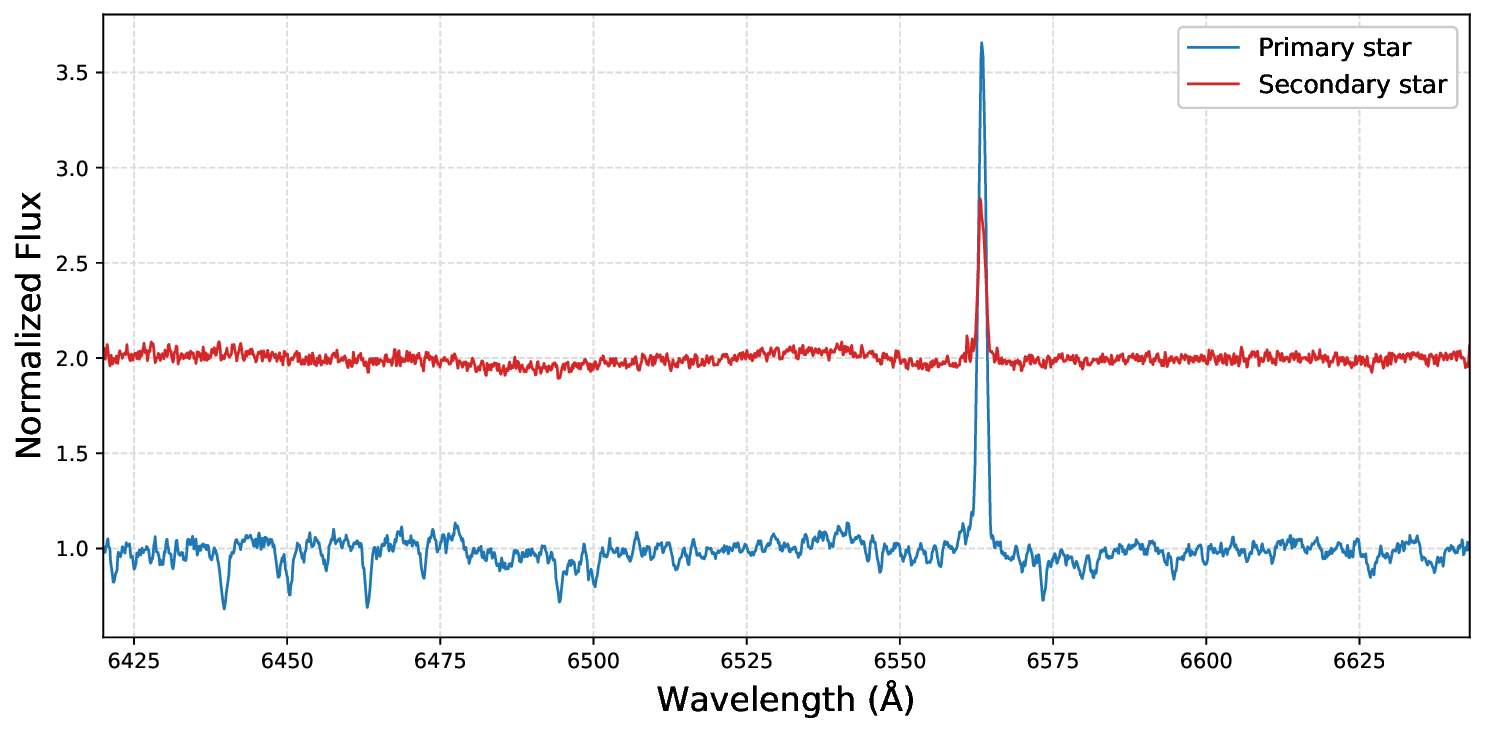}
\caption{Disentangled spectra of the primary (blue line) and secondary (red line) components around the $H\alpha$ region. For clarity, the spectrum of the secondary star has been shifted upward by 1.0 in normalized flux.}
\label{disentangled_spectra}
\end{figure}

The disentangled spectra of the primary and secondary stars are displayed in Fig.~\ref{disentangled_spectra}. The Stellar atmospheric parameters of the primary star were derived using the spectral analysis package iSpec\citep{2014A&A...569A.111B} and the input spectrum of the primary component, obtained from spectral disentangling. To prevent magnetic activity contamination, the region containing the H$\alpha$ feature was masked prior to parameter optimization. Parameter determination was carried out by interpolating within the pre-computed PHOENIX BT-Settl synthetic grid\citep{2014IAUS..299..271A}. $T_{\text{eff}}$, $\log g$, and $[\text{M/H}]$ were optimized as free parameters by minimizing the reduced $\chi^2$ over pre-selected line regions, while the microturbulence velocity $v_{\text{mic}}$ and $\alpha$-enhancement were fixed according to the pre-computed model grid. Finally, the stellar atmospheric parameters of the primary star were determined to be $T_{\text{eff}}$ = $3866^{+118}_{-118}$, $\log g$ = $4.63^{+0.41}_{-0.41}$, and $[\text{M/H}]$ = $-0.23^{+0.20}_{-0.20}$. Due to the low flux contribution of the secondary star, its disentangled spectrum exhibits a relatively low signal-to-noise ratio and shallow absorption features. Consequently, we were unable to derive reliable atmospheric parameters for the secondary component from spectral synthesis fitting.


\begin{table}[htbp]
\centering
\caption{Radial velocities measured from the spectral disentangling method.}
\label{tab:rv_phase_spd}
\begin{tabular}{cccccc}
\hline\hline
BJD       & Phase    & RV1        & $(O - C)_1$ & RV2        & $(O - C)_2$ \\
2450000+  &          & km/s       & km/s        & km/s       & km/s    \\
\hline
8451.1662 & 0.8598   &  41.9908   & 0.1148      & -71.5039   & 0.1175 \\
8451.1822 & 0.8590   &  40.8787   & 0.0318      & -68.3267   & 1.1537 \\
8451.2148 & 0.8715   &  37.9399   & 0.1486      & -65.3470   & 0.5588 \\
8512.0290 & 0.1902   & -49.4254   & 0.2416      &  84.3411   & 0.6797 \\
8512.0449 & 0.1963   & -48.8673   & 1.0222      &  85.0779   & 0.2161 \\
8512.0609 & 0.2024   & -50.4190   & 0.1023      &  86.6316   & 0.6950 \\
8512.0776 & 0.2088   & -51.2796   & 0.1788      &  86.9834   & 0.0612 \\
9187.1989 & 0.0782   & -25.2021   & 0.1877      &  42.5042   & 0.0451 \\
9187.2155 & 0.0846   & -26.2424   & 0.6203      &  47.4912   & 1.7980 \\
9187.2315 & 0.0907   & -28.3682   & 0.2245      &  51.8207   & 3.1847 \\
9210.1220 & 0.8679   &  38.8469   & 0.0787      & -65.6508   & 0.5615 \\
9210.1386 & 0.8742   &  36.8958   & 0.5651      & -63.3040   & 0.4167 \\
9217.1300 & 0.5550   &  18.1051   & 0.1048      & -30.4320   & 0.1863 \\
9626.9557 & 0.6991   &  51.5954   & 1.4109      & -84.2445   & 1.1190 \\
9626.9709 & 0.7049   &  50.2589   & 0.4990      & -86.2235   & 0.1155 \\
9626.9862 & 0.7108   &  51.6434   & 0.3804      & -86.8593   & 0.3388 \\
\hline\hline  
\end{tabular}
\end{table}

\subsection{Radial velocities measured with MESS/TODCOR}
\label{subsec:q_todcor}

As an independent measurement of both components, we analysed all 53 individual LAMOST-MRS red-arm exposures with the Multi-Epoch Spectroscopic Solver (\emph{MESS}; \citealt{2026MNRAS.546.stag250N}) and TODCOR \citep{1994ApJ...420..806Z}. Standard \emph{MESS} jointly optimizes two stellar templates over all available epochs. Because the signal-to-noise ratios of the present spectra were too low to construct stable stellar templates, the two H$\alpha$ emission components were instead represented by unit-height Gaussian templates, $g_j(v)=\exp[-v^2/(2\sigma_j^2)]$. Only the widths $\sigma_1$ and $\sigma_2$ were shared among epochs; the component amplitudes and linear background were fitted independently for each exposure.

For each spectrum, a linear continuum was fitted to the line-free intervals 6540--6550 and 6580--6590~\AA{}, and the normalized 6550--6580~\AA{} region was resampled on a constant-velocity log-wavelength grid with a spacing of $4~\mathrm{km\,s^{-1}}$. Narrow outlying pixels were masked, and preliminary one- and two-Gaussian fits supplied starting velocities and quality diagnostics. The Gaussian widths were then optimized simultaneously over the eligible epochs by maximizing the \emph{MESS} combined-correlation statistic
\begin{equation}
 w_m={\rm SNR}_m^2\,\operatorname{Var}\!\left(f_m-\langle f_m\rangle\right),
 \qquad
 S^2=\frac{\sum_m w_m\left(R_m^\star\right)^2}{\sum_m w_m},
 \label{eq:mess_halpha_score}
\end{equation}
where $R_m^\star$ is the maximum TODCOR correlation for epoch $m$. An $11\times11$ grid spanning $10\leq\sigma_{1,2}\leq100~\mathrm{km\,s^{-1}}$ served as the anchor for a subsequent exhaustive, localized multi-parameter optimization at each node, followed by the rejection of low-correlation training outliers. This multi-epoch optimization is the principal advantage of \emph{MESS}: one common template pair is constrained by all informative spectra, with each epoch weighted by both its SNR and H$\alpha$ line information.

For the optimized templates, TODCOR evaluates the two-dimensional correlation \citep{1994ApJ...420..806Z}
\begin{equation}
 R_{\alpha}(v_1,v_2)=
 \frac{C_1(v_1)+\alpha C_2(v_2)}
 {\left[1+2\alpha C_{12}(v_2-v_1)+\alpha^2\right]^{1/2}},
 \label{eq:todcor_halpha}
\end{equation}
where $C_1$ and $C_2$ correlate the observed profile with the two templates and $C_{12}$ is their mutual correlation. Its joint maximum gives the two velocities at each epoch, reducing blending and component cross-talk relative to separate one-dimensional measurements. The velocities were refined with a two-template profile-likelihood fit, and the $1\sigma$ errors were obtained from the $\Delta\chi^2=1$ intervals, including the adopted systematic floor and residual-correlation correction.

The final selection required an H$\alpha$ peak SNR of at least 2.5 and a separation of at least 0.5 times the mean FWHM. In total, 21 of the 53 spectra passed these checks. Their phases, calculated with Equation~\ref{linear-eaqution-corrected}, and measured velocities are listed in Table~\ref{RV-MESS}.

\begin{table}
\begin{center}
\caption{Measured radial velocities through H$\alpha$ emission lines with MESS/TODCOR.}
\label{RV-MESS}
\setlength{\tabcolsep}{4pt}
\begin{tabular}{llllll}
\hline\hline
Phase    & RV1    & Error  & RV2     & Error  & Observed date \\
         & (km/s) & (km/s) & (km/s)  & (km/s) &               \\\hline
0.0726   &   7.555 & 2.987 &  78.833 & 12.792 & 24 December 2021 \\
0.0882   &  -1.395 & 1.580 &  81.082 &  7.045 &  3 December 2020 \\
0.1887   & -21.386 & 1.510 & 112.327 &  4.485 & 28 January 2019  \\
0.1948   & -20.033 & 1.741 & 111.277 &  4.282 & 28 January 2019  \\
0.2009   & -21.591 & 1.521 & 116.540 &  4.778 & 28 January 2019  \\
0.2073   & -22.195 & 2.054 & 120.835 &  6.050 & 28 January 2019  \\
0.2132   & -21.832 & 2.081 & 120.839 &  7.024 & 28 January 2019  \\
0.6004   &  58.721 & 2.875 & -27.304 & 13.818 &  7 December 2020 \\
0.6068   &  64.848 & 2.539 & -27.302 & 11.276 &  7 December 2020 \\
0.6982   &  80.073 & 1.454 & -59.359 &  5.511 & 16 February 2022 \\
0.7041   &  78.862 & 1.305 & -58.456 &  5.161 & 16 February 2022 \\
0.7100   &  80.416 & 1.449 & -60.904 &  5.240 & 16 February 2022 \\
0.8443   &  72.565 & 2.179 & -48.278 &  7.550 & 28 November 2018 \\
0.8504   &  71.672 & 1.947 & -50.583 &  6.408 & 28 November 2018 \\
0.8565   &  70.403 & 1.485 & -44.551 &  5.625 & 28 November 2018 \\
0.8629   &  67.815 & 2.041 & -45.999 &  7.009 & 28 November 2018 \\
0.8655   &  68.446 & 1.235 & -43.426 &  4.606 & 26 December 2020 \\
0.8690   &  67.281 & 2.715 & -40.264 &  9.433 & 28 November 2018 \\
0.8719   &  66.301 & 1.253 & -40.367 &  5.181 & 26 December 2020 \\
0.8754   &  66.238 & 2.129 & -33.407 &  6.895 & 28 November 2018 \\
0.8816   &  63.663 & 1.844 & -45.193 &  6.425 & 28 November 2018 \\
\hline
\hline
\end{tabular}
\end{center}
\end{table}

\begin{figure}[h]
\centering
\includegraphics[width=0.8\linewidth]{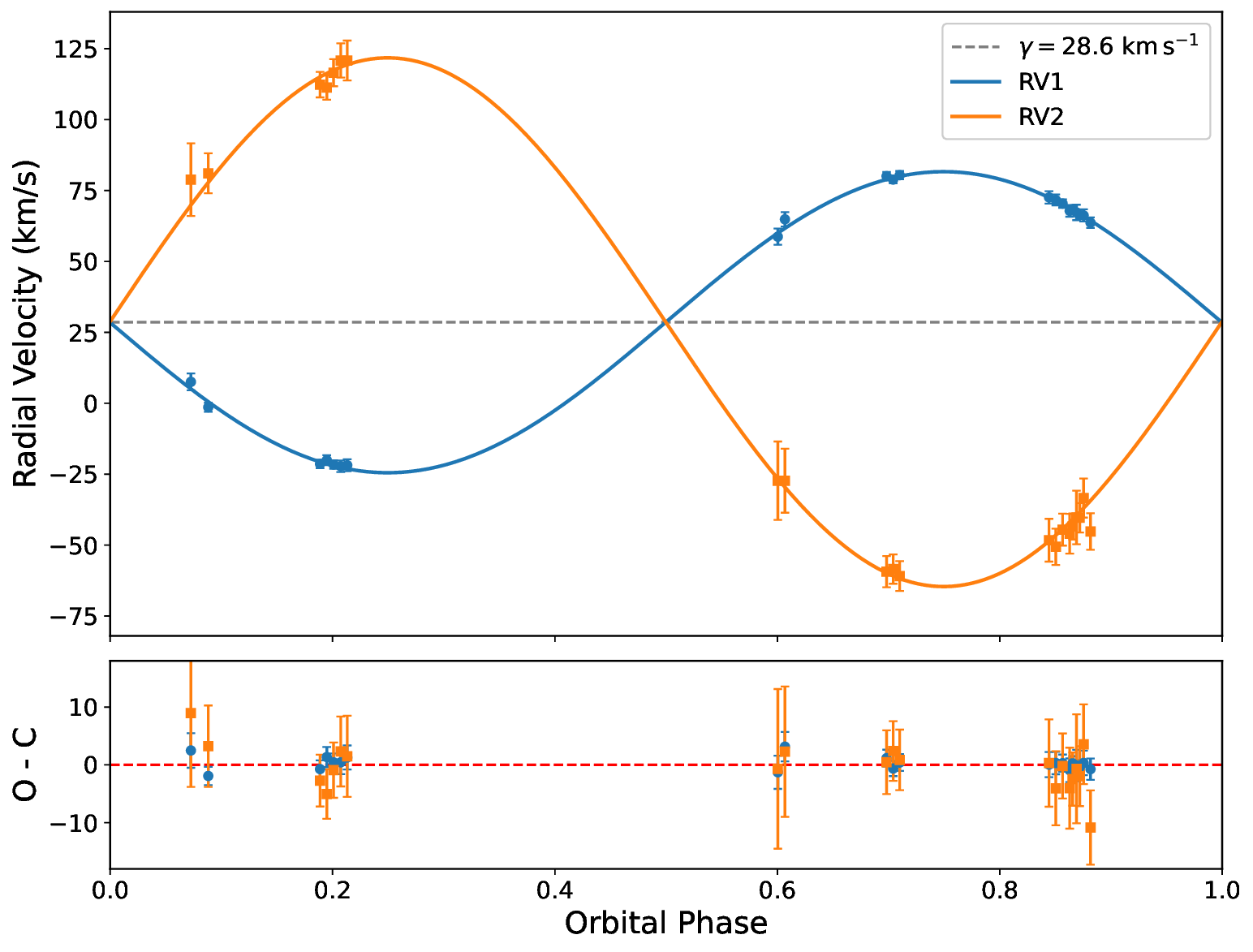}
\caption{H$\alpha$ radial velocities measured with MESS/TODCOR. The upper panel shows the 21 accepted RV1 (blue circles) and RV2 (orange squares) measurements and the best-fitting circular orbit at the fixed photometric period. The horizontal dashed line marks the systemic velocity. The lower panel shows the corresponding $O-C$ residuals; the red dashed line marks zero residual.}
\label{rv-todcor}
\end{figure}

A Wilson fit \citep{1941ApJ....93...29W} to the 21 velocity pairs yielded $\gamma=28.303\pm0.335~\mathrm{km\,s^{-1}}$ and $q=0.56926\pm0.00666$. A circular SB2 orbit fitted at the fixed photometric period $P=2.6079614$~d gave $\gamma=28.580\pm0.231~\mathrm{km\,s^{-1}}$, $K_1=53.070\pm0.290~\mathrm{km\,s^{-1}}$, $K_2=93.201\pm0.931~\mathrm{km\,s^{-1}}$, $q=K_1/K_2=0.56942\pm0.00653$, $T_0=58453.005$ MJD, and $\chi^2_{\rm red}=0.371$. The phase-folded solution and residuals are shown in Fig.~\ref{rv-todcor}. The minimum mass of the primary and secondary stars are calculated to be $M_1 \sin^3 i$ = $0.539^{+0.012}_{-0.012}$ $M_\odot$ and $M_2 \sin^3 i$ = $0.307^{+0.005}_{-0.005}$ $M_\odot$.

The MESS/TODCOR mass ratio agrees with the direct H$\alpha$ result, $q=0.576\pm0.005$, within $1\sigma$, and with the spectral-disentangling result, $q=0.586\pm0.004$, at approximately $2.2\sigma$. The value of $K_1$ is consistent with both independent analyses, whereas $K_2$ is $2.0$--$3.2~\mathrm{km\,s^{-1}}$ higher. All methods therefore support a low mass ratio of $q\simeq0.57$--$0.59$.

\section{Inversion of light curves and fundamental parameters}
\label{sect:lc}

\begin{figure}[h]
\centering
\includegraphics[width=0.8\linewidth]{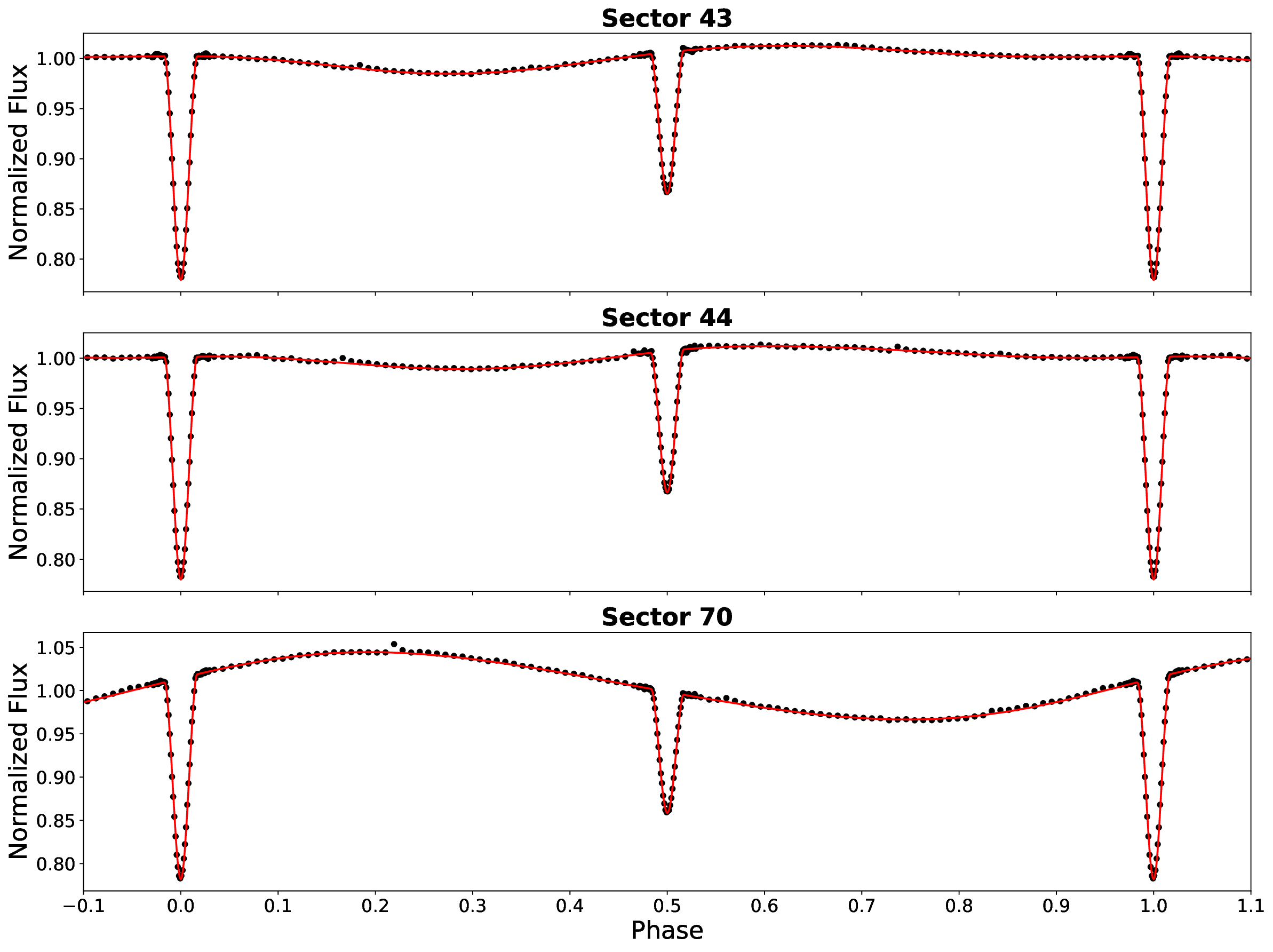}
\caption{Phase-folded light curves of TIC 349005337 for TESS Sectors 43, 44, and 70. The black dots represent the binned observational data, while the red solid lines denote the best-fit theoretical models. The phase coverage is extended from -0.1 to 1.1 to provide a complete view of the eclipsing events.}
\label{lcs-fit}
\end{figure}

To derive the fundamental parameters of the binary system, the Wilson–Devinney (WD) program \citep{1971ApJ...166..605W} was used to fit the light curves shown in Fig.~\ref{tess-lc}. Prior to the fitting, the data from each sector were converted into orbital phase and binned into 200 data points. Among them, 50 points were assigned to the primary eclipse and 50 points to the secondary eclipse, covering the phase ranges from $-0.03$ to $0.03$ and from $0.47$ to $0.53$, respectively. The remaining 100 points were assigned to outside the eclipses. The binned light curves are displayed in Fig. \ref{lcs-fit} with black dots. During the fitting process, Mode 2 for detached configuration with both components well within their Roche lobes is adopted. Although the atmospheric parameters from the LAMOST MRS pipeline provide an estimate of the effective temperature, the automated fitting relies on single-star atmospheric models. As demonstrated by \citet{2020ApJ...898...47F}, fitting a composite binary spectrum with a single-star model introduces systematic biases away from the primary star's true effective temperature. To mitigate this effect, we adopted the temperature derived from spectral disentangling, which isolates the primary component's spectrum and yields a robust estimate of $T_{\text{1}}$. Thus, the effective temperature of the primary star was assumed to be $T_1$ = $3866 K$ according to the measurements in Section~\ref{subsec:q_spd}, and adopted it for modeling the light curves of TIC 349005337. Considering the effective temperature of the primary star, a convective common envelope was assumed. Thus, the bolometric albedo coefficients of the primary and secondary stars are assumed to be $A_1$ = 0.5 and $A_2$ = 0.5 \citep{1969AcA....19..245R}. The gravity-darkening exponents of the two component stars are assumed to be $g_1$ = 0.32 and $g_2$ = 0.32 \citep{1967ZA.....65...89L}. The limb-darkening coefficients were chosen from \citet{1993AJ....106.2096V}'s tables according to the relevant stellar parameters. The mass ratio of TIC 349005337 was obtained via three methods in Section~\ref{sect:lamost}, and its mean value of $q = 0.577$ was fixed during the light curve fitting. In the WD modeling, the free parameters included the orbital inclination of the binary system ($i$), the surface effective temperature of the secondary component ($T_2$), the surface potential of the primary star ($\Omega_1$), the surface potential of the secondary star ($\Omega_2$), and the luminosity of the primary component ($L_1$). To account for asymmetries in the observed light curve shown in Fig. \ref{lcs-fit}, a cool star spot was included on each of the primary and secondary components. As the angular radius ($r$) and temperature factor ($T_f$) of a star spot are correlated in the light curve fitting, the temperature factors were fixed to be $T_f$ = 0.72 to break the degeneracy between starspot temperature and coverage fraction in the fitting, which is in line with typical starspot contrast properties observed in active late-type stars (\citealp{2005LRSP....2....8B}; \citealp{2009A&ARv..17..251S}). Then, the free parameters for spots are their latitude ($\theta$), longitude ($\phi$) and angular radius ($r$). 

The final parameters determined by the WD program are presented in Table \ref{WD_results}. As the binary system's parameters derived from the three sectors showed slight discrepancies, their weighted averages are calculated as the final results, which are listed in the last column of Table \ref{WD_results}. The surface effective temperature of the secondary component is determined to be $T_2$ = 3619($\pm4$)K, revealing that it is also an M-type star. The equality of the four directional radii ($r_{pole}$, $r_{side}$, $r_{back}$, and $r_{point}$) indicates that both of the primary and secondary stars are spherically symmetric, suggesting that the two components haven't been tidally distorted. Combining the orbital inclination of the binary system ($i = 86.76^\circ \pm 0.03^\circ$) and the minimum mass determined in Section \ref{sect:lamost}, the masses of the primary and secondary stars are calculated to be $M_1 = 0.517(\pm 0.008)M_\odot$ and $M_2 = 0.298(\pm 0.004)M_\odot$, respectively. According to Kepler's Third Law:
\begin{equation}
\frac{a^3}{P^2} = \frac{G(M_1 + M_2)}{4\pi^2}, \label{Kepler2}
\end{equation}
or
\begin{equation}
a = \frac{P(K_1 + K_2)}{2\pi \sin i},
\end{equation}
the semi-major axis of the binary system is calculated to be $a = 7.44(\pm 0.04)R_\odot$. The radii of the two component stars can be calculated with:
\begin{equation}
R = r \times a, \label{radius}
\end{equation}
where $r_1 = 0.0696(\pm0.0003)$ and $r_2 = 0.0479(\pm0.0004)$ since both components exhibit spherical geometry, indicating negligible tidal distortion. Therefore, the radius of the primary star is $R_1 = 0.518(\pm 0.003)R_\odot$, and the radius of the secondary star is $R_2 = 0.357(\pm 0.003)R_\odot$. The luminosities of the components were derived using the Stefan-Boltzmann law, 
\begin{equation}
L = 4\pi R^2 \sigma T_{eff}^4, \label{lum}
\end{equation}
based on the calculated radii and effective temperatures. Thus, the luminosities are derived to be $L_1 = 0.054(\pm 0.007)L_\odot$, and $L_2 = 0.020(\pm 0.001)L_\odot$. The surface gravities ($\log g$) of the primary and secondary components were derived from Newton's law of universal gravitation, expressed as:
\begin{equation}
    g = \frac{GM}{R^2},\label{logg}
\end{equation}
where $M$ and $R$ are the absolute mass and radius of each star, and $G$ denotes the universal gravitational constant. The resulting values are $\log g_1 = 4.722(\pm 0.005)$ for the primary star and $\log g_2 = 4.808(\pm 0.008)$ for the secondary star. All fundamental parameters are tabulated in Table \ref{physical_parameters}.

\begin{table}
\begin{center}
\caption{Photometric solutions for TIC 349005337}\label{WD_results}
\setlength{\tabcolsep}{4pt}
\begin{tabular}{lllll}
\hline\hline
Parameters                 &   Sector 43              &   Sector 44                &   Sector 70             &     Weighted Mean           \\\hline
$T_{1}(K)   $              &  3866(fixed)             &  3866(fixed)               &  3866(fixed)            &     3866(fixed)              \\
q ($M_2/M_1$ )             &  0.577(fixed)            &  0.577(fixed)              &  0.577(fixed)           &     0.577(fixed)         \\
$i(^{\circ})$              &  86.75($\pm0.02$)        &  86.76($\pm0.03$)          &  86.78($\pm0.03$)       &     86.76($\pm0.03$)            \\
$\Omega_{in}$              &  3.02                    &  3.02                      &  3.02                   &     3.02                     \\
$\Omega_{1}$               &  14.96($\pm0.04$)        &  14.93($\pm0.05$)          &  14.97($\pm0.08$)       &     14.95($\pm0.06$)        \\
$\Omega_{2}$               &  13.17($\pm0.08$)        &  13.26($\pm0.11$)          &  13.36($\pm0.14$)       &     13.26($\pm0.11$)         \\
$T_{2}(K)$                 &  3619($\pm2$)            &  3622($\pm2$)              &  3615($\pm7$)           &     3619($\pm4$)              \\
$\Delta T(K)$              &  247($\pm2$)             &  244($\pm2$)               &  251($\pm7$)            &     247($\pm4$)                 \\
$L_{1}/(L_{1}+L_{2}$)      &  0.771($\pm0.001$)       &  0.773($\pm0.001$)         &  0.777($\pm0.002$)      &     0.774($\pm0.001$)       \\
$r_{1}(pole)$              &  0.0695($\pm0.0002$)     &  0.0697($\pm0.0002$)       &  0.0694($\pm0.0004$)    &     0.0695($\pm0.0003$)     \\
$r_{1}(point)$             &  0.0695($\pm0.0002$)     &  0.0697($\pm0.0002$)       &  0.0695($\pm0.0004$)    &     0.0696($\pm0.0003$)     \\
$r_{1}(side)$              &  0.0695($\pm0.0002$)     &  0.0697($\pm0.0002$)       &  0.0695($\pm0.0004$)    &     0.0696($\pm0.0003$)     \\
$r_{1}(back)$              &  0.0695($\pm0.0002$)     &  0.0697($\pm0.0002$)       &  0.0695($\pm0.0004$)    &     0.0696($\pm0.0003$)     \\
$r_{2}(pole)$              &  0.0483($\pm0.0003$)     &  0.0478($\pm0.0004$)       &  0.0476($\pm0.0006$)    &     0.0479($\pm0.0004$)     \\
$r_{2}(point)$             &  0.0483($\pm0.0003$)     &  0.0479($\pm0.0004$)       &  0.0476($\pm0.0006$)    &     0.0479($\pm0.0004$)     \\
$r_{2}(side)$              &  0.0483($\pm0.0003$)     &  0.0479($\pm0.0004$)       &  0.0476($\pm0.0006$)    &     0.0479($\pm0.0004$)     \\
$r_{2}(back)$              &  0.0483($\pm0.0003$)     &  0.0479($\pm0.0004$)       &  0.0476($\pm0.0006$)    &     0.0479($\pm0.0004$)     \\\hline
\multicolumn{5}{c}{\textbf{Spot 1 on the primary star}}    \\
$\theta_1(^{\circ})$       &  22.1($\pm1.9$)          &  37.0($\pm4.8$)            &  13.4($\pm2.4$)         &                             \\
$\phi_1(^{\circ})$         &  260.7($\pm1.2$)         &  253.1($\pm2.6$)           &  152.7($\pm4.5$)        &                             \\
$r_1(\mathrm{rad})$        &  0.36($\pm0.02$)         &  0.23($\pm0.02$)           &  0.62($\pm0.07$)        &                            \\
$T_{f1}$                   &  0.72(fixed)             &  0.72(fixed)               &  0.72(fixed)            &                            \\\hline
\multicolumn{5}{c}{\textbf{Spot 2 on the secondary star}}   \\                                                                        
$\theta_2(^{\circ})$       &  148.4($\pm4.2$)         &  157.9($\pm4.3$)           &  146.5($\pm2.5$)        &                            \\
$\phi_2(^{\circ})$         &  221.8($\pm3.4$)         &  202.6($\pm2.9$)           &  239.0($\pm2.7$)        &                            \\
$r_2(\mathrm{rad})$        &  0.41($\pm0.05$)         &  0.52($\pm0.06$)           &  0.89($\pm0.03$)        &                            \\
$T_{f2}$                   &  0.72(fixed)             &  0.72(fixed)               &  0.72(fixed)            &                            \\
\hline
\hline
\end{tabular}
\end{center}
\end{table}

\begin{table}[ht]
\centering
\caption{Physical parameters of the primary and secondary components.}
\label{physical_parameters}
\begin{tabular}{lccc}
\toprule
Parameter                                    & Unit         & Primary (Star 1)         & Secondary (Star 2) \\
\midrule
Effective temperature ($T_{\text{eff}}$)     & K            & $3866 \pm 118$           & $3619 \pm 4$ \\
Mass ($M$)                                   & $M_\odot$    & $0.517 \pm 0.009$        & $0.298 \pm 0.004$ \\
Radius ($R$)                                 & $R_\odot$    & $0.518 \pm 0.003$        & $0.357 \pm 0.003$ \\
Luminosity ($L$)                             & $L_\odot$    & $0.054 \pm 0.007$        & $0.020 \pm 0.001$ \\
Surface gravity ($\log g$)                   & cgs          & $4.722 \pm 0.005$        & $4.808 \pm 0.008$ \\
\midrule
Semi-major axis ($a$)                        & $R_\odot$    & \multicolumn{2}{c}{$7.44 \pm 0.04$} \\
\bottomrule
\end{tabular}
\end{table}

The Gaia RUWE of TIC 349005337 is 2.281, suggesting that there may be an unresolved astrometric companion orbiting around the central eclipsing binary system \citep{2018A&A...616A...2L,2021A&A...649A...2L}. Additionally, flux contamination from known surrounding stars can be evaluated using the CROWDSAP parameters provided by the SPOC pipeline. Considering the large pixel scale of TESS ($21''$), we retrieved the CROWDSAP parameters for TIC 349005337, which are 0.998 in Sector 43, 0.991 in Sector 44, and 0.964 in Sector 70. Both the potential unresolved astrometric companion and the known neighboring stars can introduce third light ($L_3$). Since flux dilution can affect the derived physical parameters, the effect of third light was explicitly accounted for during the light curve fitting\citep{2010MNRAS.408.1689S}. To assess the sensitivity of the physical parameters to potential flux dilution, we fixed the third-light fraction to $L_3 = 5\%$ and re-fitted the light curves. The new results are $T_{2} = 3623(\pm 2)K$, $i = 86.87(\pm 0.02)^{\circ}$, $r_{1} = 0.0681(\pm0.0002)$, and $r_{2} = 0.0486(\pm0.0004)$. Then, the masses and radii of both components were recalculated as $M_{1} = 0.517(\pm0.009)M_\odot$, $M_{2} = 0.298(\pm0.004)M_\odot$, $R_{1} = 0.507(\pm0.003)R_\odot$, and $R_{2} = 0.362(\pm0.009)R_\odot$. The results indicate that assuming a $5\%$ third light fraction has almost no effect on the measured masses of the two components, whereas the radius of the primary star decreases by about $2\%$ and the radius of the secondary star increases by about $1\%$.

Since the observed light curves show distinct out-of-eclipse modulation, we included starspots in our light curve modeling and the spots' parameters are also listed in Table~\ref{WD_results}. However, the true distribution of starspots on both components cannot be uniquely constrained. To assess how different spot configurations influence the derived stellar parameters, we further incorporated four alternative spot models following the work of \citet{2011ApJ...742..123I}. In our case, a minimum of two spots is necessary to reproduce the out-of-eclipse modulations. Model 1: Two spots on the primary star; Model 2: Two spots on the secondary star; Model 3: One spot on the primary and another one on the secondary star; Model 4: Two spots on the primary and another two on the secondary star. And the newly derived parameters are listed in Table~\ref{Model_spots}. The results show that the component masses derived from different spot models are virtually identical, whereas the radii are noticeably affected, with maximum deviations reaching up to $4\%$.

\begin{table}[ht]
\centering
\caption{Sensitivity of parameters to various starspot distributions.} 
\label{Model_spots}
\begin{tabular}{lccccc}
\toprule
Parameter           & Unit         & Model 1                  & Model 2               & Model 3                  & Model 4\\
\midrule
$T_2$               & $K$          & $3616 \pm 2$             & $3619 \pm 2$          & $3619 \pm 2$             & $3624 \pm 2$\\
$i$                 & $^{\circ}$   & $86.79 \pm 0.03$         & $86.75 \pm 0.02$      & $86.75 \pm 0.02$         & $86.68 \pm 0.01$\\
$M_1$               & $M_\odot$    & $0.517 \pm 0.008$        & $0.517 \pm 0.008$     & $0.517 \pm 0.008$        & $0.518 \pm 0.008$\\
$M_2$               & $M_\odot$    & $0.299 \pm 0.004$        & $0.299 \pm 0.004$     & $0.299 \pm 0.004$        & $0.299 \pm 0.004$\\
$R_1$               & $R_\odot$    & $0.516 \pm 0.003$        & $0.518 \pm 0.003$     & $0.518 \pm 0.003$        & $0.511 \pm 0.003$\\
$R_2$               & $R_\odot$    & $0.358 \pm 0.003$        & $0.359 \pm 0.003$     & $0.360 \pm 0.003$        & $0.371 \pm 0.003$\\
\midrule
\bottomrule
\end{tabular}
\end{table}

\section{Physical properties of optical flares}
\label{sect:flares}

As shown in Fig.~\ref{tess-lc}, optical flare events occurred in TESS observations of Sector 43, Sector 44 and Sector 70. In our analysis, we explicitly retained data points flagged with quality flag 512 to preserve peak flare signals, since previous research has shown that excluding this flag can bias the measurement of flare energies \citep{2020AJ....160..219F}. The total energy released in the flare event is calculated with the following equation :
\begin{equation}
    E_{flare} = L_{star} \times \int_{T_{start}}^{T_{end}} \frac{F(t) - {F}_{quiet}}{{F}_{quiet}} dt,\label{flare_energy}
\end{equation}
where $E_{flare}$ is the total radiative energy released during the flare event, $L_{star}$ is the luminosity of the star in its ``quiet'' state (Quiescent Luminosity), $F(t)$ is the observational brightness of the star during the flare event, ${F}_{quiet}$ is the brightness of the star when no flare is occurring (Quiescent Flux), $T_{start}$ (Flare Start Time) is the moment that magnetic energy release in the stellar atmosphere begins to significantly exceed background radiation, and $T_{end}$ (Flare End Time) is the moment that the extra hot plasma has cooled and the star's brightness returns to its pre-disturbance level
\citep{1972Ap&SS..19...75G,1976ApJS...30...85L,2013ApJS..209....5S,2014ApJ...797..122D}. $\frac{F(t) - {F}_{quiet}}{{F}_{quiet}}$ is a dimensionless value representing how many times ``brighter'' the flare is compared to the star's normal state at that moment. $\int_{T_{start}}^{T_{end}} \frac{F(t) - {F}_{quiet}}{{F}_{quiet}} dt$ is measured in seconds and defined as Equivalent Duration (ED), which represents the area between the flare curve and the baseline, and tells us the flare energy is equivalent to the star ``burning'' for this many additional seconds at its quiet rate.

\begin{figure}[h]
\centering
\includegraphics[width=0.8\linewidth]{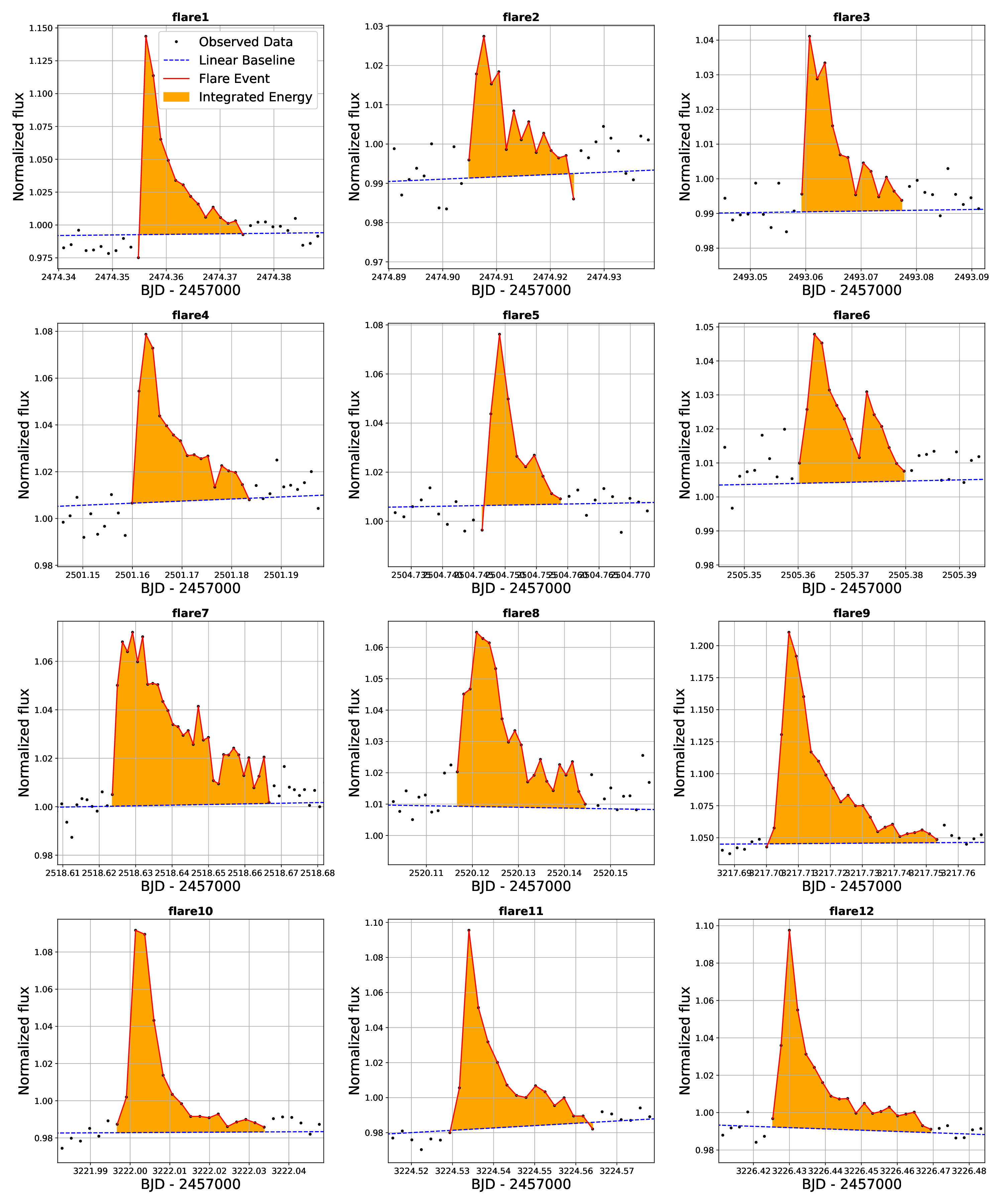}
\caption{Time-series photometry of the observed flares. The black dots represent the observed flux, the blue dashed line denotes the quiescent baseline, and the red line highlights the identified flare event. The orange shaded area represents the integrated energy (Equivalent Duration).}
\label{flares_plots}
\end{figure}

A total of 12 flare events are detected and displayed in Fig.~\ref{flares_plots}. Their properties are listed in Table \ref{tab:flares-property}. The first three columns represent the start, peak, and end times of the stellar flare, respectively. The next three columns represent the total duration, the rise-time, and the decay-time of the flare events. The rise phase of a stellar flare is an impulsive energy release driven by magnetic reconnection, where accelerated particles heat the chromosphere and trigger chromospheric evaporation, causing a rapid surge in flux. Conversely, the decay phase is a gradual cooling process where the heated plasma dissipates energy through radiative cooling and thermal conduction, accompanied by the gravitational draining of material back to the stellar surface \citep{2012Natur.485..478M,2024LRSP...21....1K}. The relative amplitude ($A_{\rm rel}$) of the flare, defined as the ratio of the peak flux increase to the star's quiescent background level ($A_{\rm rel} = \frac{F_{peak} - F_0}{F_0}$), is a key physical quantity for measuring flare intensity. It serves as a fundamental metric to characterize the flare's severity and enables a standardized comparison of eruptive activity across different stellar types. In a binary system, since the specific host of the flare event remains uncertain, we separately assume the flare occurs on each component. Accordingly, we calculate the flare's peak luminosity and the total energy released for both scenarios. In Table \ref{tab:flares-property}, $L_{\rm 1, peak}$ ($L_{\rm 2, peak}$) and $E_{\rm 1, total}$ ($E_{\rm 2, total}$) correspond to the peak luminosity and total energy released, assuming the flare event originated from the primary (secondary) component.

\begin{table*}
\centering
\caption{Parameters of the 12 detected stellar flare events}
\label{tab:flares-property}
\large
\resizebox{\textwidth}{!}{
\begin{tabular}{ccccccccccccc}
\toprule \hline
$T_{start}$  &  $T_{peak}$     &  $T_{end}$ & Duration  & Rise  & Decay  & $A_{\rm rel}$ & ED     & $L_{\rm 1, peak}$  & $E_{\rm 1, total}$ & $L_{\rm 2, peak}$  & $E_{\rm 2, total}$  \\ 
             & BJD - 2457000   &            &  (s)      &  (s)  &  (s)   &               & (s)    &  (erg s$^{-1}$)    &  (erg)              &  (erg s$^{-1}$)    &     (erg)      \\
\midrule
 2474.3548 & 2474.3562 & 2474.3743 & 1680 & 120 & 1560 & 0.152   & 70.66   & $3.14\times10^{31}$     & $1.46\times10^{34}$   & $1.16\times10^{31}$ &  $5.41\times10^{33}$       \\
 2474.9048 & 2474.9076 & 2474.9243 & 1680 & 240 & 1440 & 0.036   & 22.91   & $7.49\times10^{30}$     & $4.74\times10^{33}$   & $2.77\times10^{30}$ &  $1.75\times10^{33}$       \\
 2493.0592 & 2493.0607 & 2493.0773 & 1560 & 120 & 1440 & 0.051   & 29.24   & $1.06\times10^{31}$     & $6.04\times10^{33}$   & $3.91\times10^{30}$ &  $2.24\times10^{33}$       \\
 2501.1599 & 2501.1627 & 2501.1836 & 2040 & 240 & 1800 & 0.071   & 51.63   & $1.48\times10^{31}$     & $1.07\times10^{34}$   & $5.47\times10^{30}$ &  $3.95\times10^{33}$       \\
 2504.7463 & 2504.7491 & 2504.7588 & 1080 & 240 & 840  & 0.069   & 25.90   & $1.43\times10^{31}$     & $5.35\times10^{33}$   & $5.30\times10^{30}$ &  $1.98\times10^{33}$       \\
 2505.3603 & 2505.3630 & 2505.3797 & 1680 & 240 & 1440 & 0.086   & 32.94   & $9.00\times10^{30}$     & $6.83\times10^{33}$   & $3.33\times10^{30}$ &  $2.53\times10^{33}$       \\
 2518.6236 & 2518.6291 & 2518.6666 & 3720 & 480 & 3240 & 0.192   & 123.41  & $1.48\times10^{31}$     & $2.56\times10^{34}$   & $5.48\times10^{30}$ &  $9.47\times10^{33}$       \\
 2520.1167 & 2520.1209 & 2520.1445 & 2400 & 360 & 2040 & 0.107   & 55.90   & $1.14\times10^{31}$     & $1.16\times10^{34}$   & $4.22\times10^{30}$ &  $4.29\times10^{33}$       \\
 3217.7001 & 3217.7070 & 3217.7533 & 4600 & 600 & 4000 & 0.228   & 186.71  & $3.26\times10^{31}$     & $3.87\times10^{34}$   & $1.21\times10^{31}$ &  $1.43\times10^{34}$       \\
 3221.9967 & 3222.0014 & 3222.0338 & 3200 & 400 & 2800 & 0.129   & 85.27   & $2.29\times10^{31}$     & $1.77\times10^{34}$   & $8.48\times10^{30}$ &  $6.54\times10^{33}$       \\
 3224.5294 & 3224.5340 & 3224.5641 & 3000 & 400 & 2600 & 0.116   & 86.41   & $2.39\times10^{31}$     & $1.79\times10^{34}$   & $8.86\times10^{30}$ &  $6.62\times10^{33}$       \\
 3226.4254 & 3226.4300 & 3226.4693 & 3800 & 400 & 3400 & 0.106   & 91.00   & $2.20\times10^{31}$     & $1.88\times10^{34}$   & $8.15\times10^{30}$ &  $6.97\times10^{33}$       \\ \hline
\bottomrule            
\end{tabular}
}
\end{table*}

\section{Discussion and Conclusions}
\label{sect:conclusions}

\begin{figure}[h]
\centering
\includegraphics[width=0.8\linewidth]{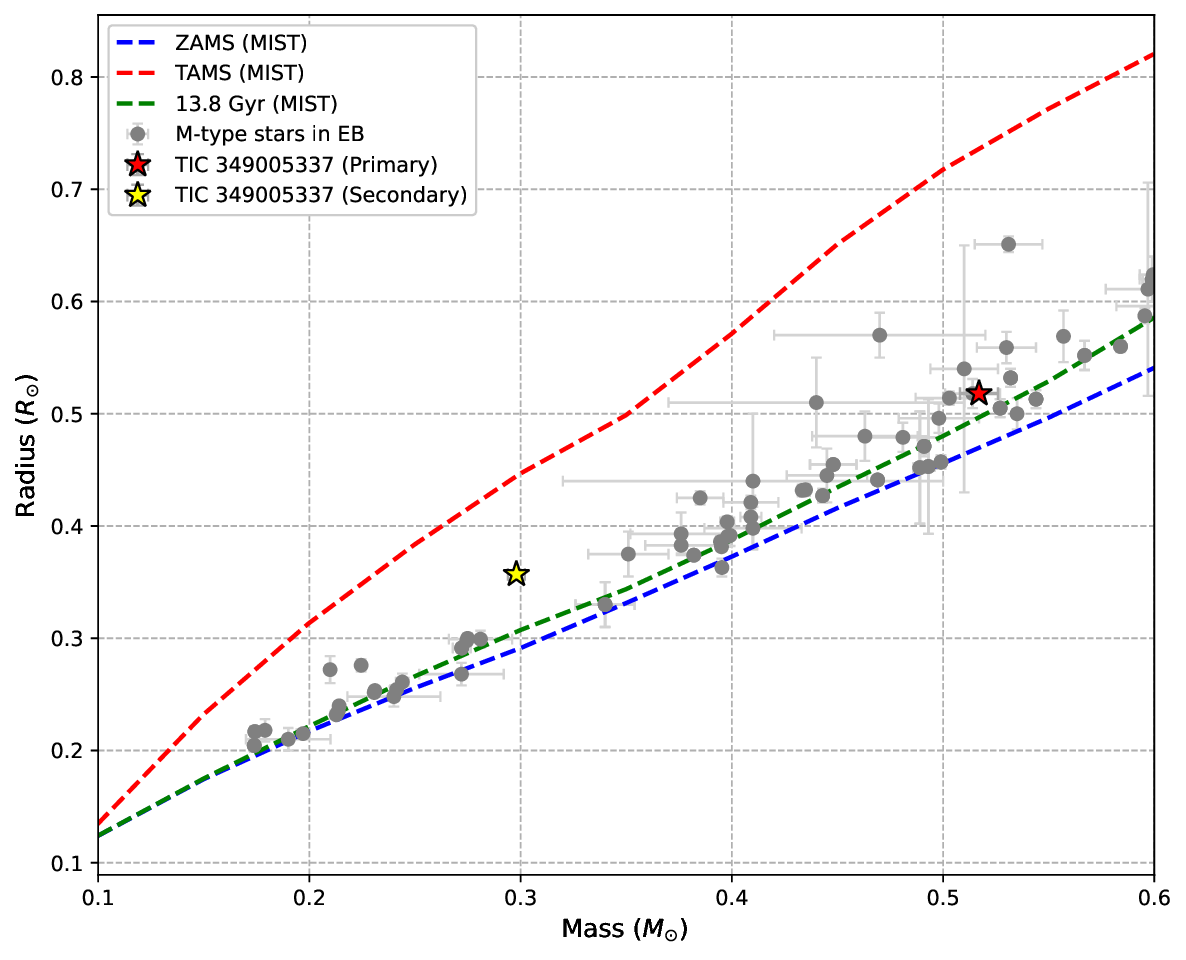}
\caption{Mass–radius relation for low-mass stars ($M \le 0.6\,M_\odot$). The primary and secondary components of TIC 349005337 are highlighted by the red and yellow stars, respectively. Grey circles with error bars represent known M-type stars in eclipsing binaries. The dashed curves correspond to the MIST theoretical models for ZAMS (blue), TAMS (red), and an age of 13.8 Gyr (green).}
\label{M-R_relation}
\end{figure}

M-dwarf stars have extremely long evolutionary lifetimes and span the critical mass threshold for full convection, making them natural laboratories for studying the internal physical processes of low-mass stars and magnetic activity mechanisms. Precise determination of their physical parameters is of great significance for directly testing and calibrating low-mass stellar evolutionary models. To evaluate the evolutionary state and physical properties of TIC 349005337, we plotted its primary and secondary components on the mass–radius ($M$–$R$) diagram alongside a benchmark sample of M-type stars in eclipsing binaries collected from \citet{2012MNRAS.426.1507B}, \citet{2018MNRAS.481.1083P} and \citet{2024MNRAS.531.1765L}, as shown in Fig.~\ref{M-R_relation}. And also, we adopted the stellar evolutionary tracks from the MESA Isochrones and Stellar Tracks (MIST; \citealp{2016ApJ...823..102C,2016ApJS..222....8D}), which are calculated using the Modules for Experiments in Stellar Astrophysics (MESA; \citealp{2011ApJS..192....3P,2013ApJS..208....4P}). The results indicate that, compared with theoretical evolutionary tracks, the measured parameters of M dwarfs generally exhibit a radius inflation effect. For the case of TIC 349005337, we accounted for the effects of third light and starspot distribution. Although they have a certain impact on the measured radius, it is not sufficient to compensate for the observed radius inflation effect. To more clearly elucidate this discrepancy between theory and observations, precise observations and detailed analyses of more M dwarfs are highly necessary in future studies.

During the $T_{\text{obs}} = \text{65.1}$ days of TESS observations, we detected 12 distinct flare events, corresponding to an average flare occurrence rate of $\nu \approx \text{0.18} \text{ flares/day}$ for the binary system, and the average per-star flare occurrence rate is $\nu_{\text{per-star}} \approx 0.09 \text{ day}^{-1}\text{star}^{-1}$. This rate is substantially higher than $74\%$ of the single low mass M stars from \citet{2022ApJ...935..104M}. Coupled with the total energy release across these events, the quantitative evidence confirms that the system maintains a high state of magnetic activity. The most energetic flare event, occurring at BJD 2460217.7001 with an energy of $3.87 \times 10^{34}$ erg (for Star 1), highlights the efficiency of magnetic reconnection in this system. The secondary component ($M_2 = 0.299 M_\odot$) falls below the fully convective boundary ($\sim 0.35 M_\odot$). In such a star, the absence of a tachocline implies that the magnetic field is generated by a distributed dynamo different from that of a solar-type star. The rapid succession of flares suggests a persistent re-configuration of the magnetic field lines between the two components, likely enhanced by tidal interactions in this close binary system. The combined Equivalent Duration (ED) and high peak luminosities indicate a harsh radiation environment. For any potential planetary bodies orbiting this system, such frequent superflares would pose a significant threat to atmospheric stability, potentially leading to atmospheric stripping or profound alterations in photochemical evolution \citep{2016PhR...663....1S}.

In summary, we performed a comprehensive analysis of the M-type detached eclipsing binary TIC 349005337 using TESS photometry and LAMOST medium-resolution spectroscopy. The precise physical parameters for both components are determined. The system consists of an M-dwarf primary ($0.517 \pm 0.008 M_\odot$) and a fully convective secondary ($0.299 \pm 0.004 M_\odot$). Both stars exhibit significant radius inflation, that their radii appear larger than predicted by standard models, suggesting complex interactions between magnetic activity and stellar structure. A total of 12 flare events with energies up to $3.87 \times 10^{34}$ erg are identified. The flare morphology, characterized by rapid rise and gradual decay phases, aligns with the magnetic reconnection theory. The high flare frequency confirms that TIC 349005337 is a magnetically active M-dwarf binary observed in the TESS sectors. The presence of such energetic flares on a fully convective star ($M < 0.35 M_\odot$) offers critical constraints for magnetic dynamo models in the low-mass regime. TIC 349005337 may serve as an ideal laboratory for studying the interplay between binary dynamics, magnetic field generation, and the resulting atmospheric phenomena in M-dwarfs. Its X-ray radiation has also been detected by the ROSAT satellite, derived a logarithm of X-ray to bolometric luminosity to be $\log\left(\frac{L_{\mathrm{X}}}{L_{\mathrm{bol}}}\right)$ = -3.02, which also confirmed that it was in an active stellar state \citep{2011ApJ...743...48W}. Future multi-wavelength observations will be essential to further probe the coronal properties and the long-term magnetic evolution of this highly active system.

\begin{acknowledgments}
We thank the anonymous reviewer for the invaluable feedback and constructive suggestions, which significantly enhanced the quality and rigor of this manuscript.
This work is supported by the International Partnership Program of Chinese Academy of Sciences (No. 020GJHZ2023030GC), the Young Talent Project of Yunnan Revitalization Talent Support Program, and the Yunnan Fundamental Research Projects (Grant Nos. 202401AW070004, 202503AP140013, 202401AS070046, 202601AT070222). This research has made use of the SIMBAD database, operated at CDS, Strasbourg, France. This paper has made use of data collected by the Transiting Exoplanet Survey Satellite (TESS) mission, for which funding is provided by the NASA Explorer Program. All the {\it TESS} data used in this paper can be found in MAST \citep{https://doi.org/10.17909/t9-nmc8-f686}. This work made use of the data from LAMOST (Large Sky Area Multi-Object Fiber Spectroscopic Telescope, also known as the Guoshoujing Telescope) (https://cstr.cn/31118.02.LAMOST). LAMOST is a Chinese national mega-science facility, operated by National Astronomical Observatories, Chinese Academy of Sciences.
\end{acknowledgments}

\bibliography{references}{}
\bibliographystyle{aasjournal}



\end{document}